# Subwavelength topological interface modes in a multilayered vibroacoustic metamaterial

**Majdi O. Gzal[1*], Joshua R. Tempelman[2], Kathryn H. Matlack[1], Alexander F. Vakakis[1]**

[1] Department of Mechanical Science and Engineering, University of Illinois, Urbana, USA
[2] Los Alamos National Laboratory, Los Alamos, USA

**\* Correspondence:**
Majdi O. Gzal
gzal@illinois.edu

## Abstract

We present a systematic and rigorous analytical approach, based on the transfer matrix methodology, to study the existence, evolution, and robustness of subwavelength topological interface states in practical multilayered vibroacoustic phononic lattices. These lattices, composed of membrane-air cavity unit cells, exhibit complex band structures with various bandgaps, including Bragg, band-splitting induced, local resonance, and plasma bandgaps. Focusing on the challenging low-frequency range and assuming axisymmetric modes, we show that topological interface states are confined to Bragg-like band-splitting induced bandgaps. Unlike the Su-Schrieffer-Heeger model, the vibroacoustic lattice exhibits diverse topological phase transitions across infinite bands, enabling broadband, multi-frequency vibroacoustics in the subwavelength regime. We establish three criteria for the existence of these states: the Zak phase, surface impedance, and a new reflection coefficient concept, all derived from transfer matrix components. Notably, we provide an explicit expression for the exact location of topological interface states within the band structure, offering insight for their predictive implementation. We confirm the robustness of these states against structural variations and identify delocalization as bandgaps narrow. Our work provides a complete and exact analytical characterization of topological interface states, demonstrating the effectiveness of the transfer matrix method. Beyond its analytical depth, our approach provides a useful framework and design tool for topological phononic lattices, advancing applications such as efficient sound filters, waveguides, noise control, and acoustic sensors in the subwavelength regime. Its versatility extends beyond the vibroacoustic systems, encompassing a broader range of phononic and photonic crystals with repetitive inversion-symmetric unit cells.

# 1. Introduction

Topological interface modes have emerged as a prominent research focus in topological physics, originating from advances in condensed matter physics that led to the discovery of topological insulators, which leverage various band phenomena, including the anomalous Hall effect, valley Hall effect, and edge modes [1-5]. These modes, often associated with robust states localized at the boundaries or interfaces between distinct topological phases, exhibit remarkable properties that are insensitive to local perturbations, making them highly desirable for various applications, including waveguiding, energy transfer, and information processing. Due to these unique advantages, the concept of topologically protected interface modes has been extended to various classical areas of physics, including electromagnetic waves in photonic crystals [6-13], as well as acoustic [14-28] and elastic [29-36] waves in phononic crystals (PCs).

The realization of topological interface modes relies in the topological characteristics of materials and systems, typically quantified by topological invariants such as the Chern number in two-dimensional (2D) systems or the Zak phase in one-dimensional (1D) counterparts. Central to these modes is the concept of bulk-boundary correspondence, which asserts that nontrivial topological phases in a material's bulk inherently generate protected states at its boundaries or interfaces with regions exhibiting different topological properties [37,37].

The exploration of 1D topological systems can be traced back to the seminal works on the Su-Schrieffer-Heeger (SSH) model, originally proposed to describe the electronic properties of polyacetylene chains [39,40]. The widely studied SSH model, characterized by alternating hopping amplitudes in a 1D chain, is a prototypical system for studying topological phases in 1D. This model demonstrated the concept of topological invariants, specifically, the Zak phase, which is related to the variation of the geometric phase over a Brillouin zone [41]. Indeed, the Zak phase classifies the topology of bulk bands, and is crucial for predicting the existence of topologically protected interface states—or localized modes that form at the boundaries between regions with distinct Zak phases. In 1D systems, a topological phase transition is related to a sudden change in a topological invariant, such as the Zak phase, caused by variations in system parameters like material composition or structural geometry. In the bulk, this transition occurs when an existing band gap closes and reopens for changing system parameters, leading to fundamental changes to the system's topological properties. These transitions govern the creation or annihilation of protected interface states, enabling precise control over wave confinement and guiding. As a result, the Zak phase becomes a useful tool for the design of topological electronic, photonic, and phononic systems.

Generalizations of the SSH model often reveal nontrivial and intriguing phenomena, enhancing our understanding of topological phases. For instance, unconventional topological phases of polaritons in a cavity waveguide, modeled as an extended SSH model, have been reported to demonstrate how strong light-matter coupling can lead to a breakdown of the bulk-edge correspondence, both theoretically [42] and experimentally [43]. Moreover, owing to its relative simplicity, the SSH model has been successfully implemented in diverse physical settings, such as in acoustic [16] and electromagnetic [44] systems. However, these topological states, typically arising within Bragg bandgaps, are generally restricted to high-frequency wave modes. This presents a significant challenge for acoustic waves in the audible range, which require bulky structures to achieve low-frequency topological states due to their relatively large wavelengths. In [24], topological interface states in subwavelength acoustic systems were explored using a spring-mass discrete model, but this approach limits the practical applicability to real-world sound waves.

In the present study, we focus on the topological interface states in phononic systems, with a particular emphasis on membrane-type vibroacoustic metamaterials, a less explored area that holds promising potential for low-frequency sound tailoring and manipulation [45,46]. To simultaneously achieve broadband low-frequency topological interface states within the subwavelength scale, we propose a vibroacoustic phononic metamaterial composed of multilayered membrane-air cavity units. Extending the topological interface states concept to vibroacoustic systems is particularly challenging due to their complex band structures, characterized by multiple interacting bandgaps. This challenge was recently addressed by the authors, by performing a thorough analytical characterization of the band structures of vibroacoustic phononic metamaterials composed of repeated monolayered [47] and multilayered [48] membrane-air cavity unit cells, based on an analytical method initially introduced in [49]. This detailed analysis of the vibroacoustic multilayered phononic lattice provided the essential groundwork for the current study, enabling a purely analytical investigation of topological interface states in practical vibroacoustic phononic metamaterials.

A distinctive feature of the multilayered vibroacoustic metamaterial considered herein is that, by adjusting the properties of the membrane, sound-membrane interactions can be tuned to achieve broadband sub-Bragg bandgaps, termed band-splitting induced bandgaps, within the subwavelength scale [48]. Unlike local resonance bandgaps, such subwavelength bandgaps are broadband and exhibit Bragg-like attenuation at frequencies well below the first-order Bragg diffraction, rendering them promising candidates for supporting low-frequency topological interface states.

Moreover, while the Zak phase is a well-established topological invariant used to identify interface states in various SSH model extensions, its abstract nature complicates the observation and interpretation of phase transitions and the role of different physical parameters, particularly when dealing with complex systems beyond artificial mass-spring models, such as the proposed vibroacoustic system. Therefore, in this study, we complement the Zak phase analysis with the concept of surface impedance, as proposed in [44], to derive explicit criteria for the existence and precise location of interface states within the topological bandgap. This approach relies on components of the transfer matrix that are directly related to the system's physical parameters. Furthermore, we introduce a new approach to characterize interface states using the reflection coefficients of semi-infinite, topologically distinct lattices. Hence, the presented analysis provides fresh insights into the nontrivial topological properties of low-frequency, subwavelength topological interface states in vibroacoustic phononic metamaterials with inversion-symmetric unit-cells. Overall, this study, grounded solely in the transfer matrix formalism, is driven by the aim of advancing the practical design of topological modes. It offers both physical insights and direct design tools for constructing topological interface states at specific target frequencies within vibro-acoustic lattices. While the effectiveness of the transfer matrix method for analyzing edge states has been highlighted in recent work [50], our approach presents a significantly different perspective.

This work is organized as follows. Section 2 briefly addresses the sound-membrane interaction problem within a unit cell featuring inversion symmetry in a multilayered vibroacoustic lattice of infinite extent. We present exact analytical solutions based on Bloch's theorem and the transfer matrix method, enabling a detailed characterization of the band structure of these phononic metamaterials. In Sections 3 and 4, we present an analytical investigation of the existence, location, and robustness of topological interface states using the transfer matrix formalism. This analysis is conducted from two perspectives: in Section 3, we focus on the infinite lattices, utilizing concepts such as the Zak phase, surface impedance, and reflection coefficients; in Section 4, we examine the finite composite lattice through scattering spectra and eigenfrequency analysis. Section 5 applies these methods to demonstrate the formation, robustness, and evolution of interface states. Finally, Section 6 provides some concluding remarks with a discussion of the implications and potential technological applications of our findings.

## 2. Vibroacoustic phononic metamaterial of infinite extent: analysis and characterization

In this section, we begin with a detailed model description, outlining the key components of the coupled sound-membrane vibroacoustic system. Then, we establish the governing equations of motion and boundary conditions. Thereafter, an analytical exact solution for the coupled system is derived, providing a framework to describe the vibroacoustic interactions. Next, we employ the transfer matrix formalism to systematically analyze wave propagation across the multilayered structure. This is followed by a derivation of local impedance, scattering matrices, and reflection-transmission spectra to characterize wave behavior at interfaces. Finally, we conduct a complete band structure characterization, offering insights into the dispersion properties and the emergence of bandgaps in the system.

### 2.1. Model description

We investigate a cylindrical (axisymmetric) acoustic duct embedded with an infinite array of multilayered membrane resonators, forming a phononic vibroacoustic metamaterial. This structure consists of coupled multilayered repetitive unit cells creating a periodic lattice. Figure 1(a) provides a schematic of three adjacent unit cells in this periodic arrangement which is of infinite spatial extent. Each multilayered unit cell includes layers of membrane-air cavity resonators arranged in the following sequence: Half of Cavity 2, Membrane 1, Cavity 1, Membrane 2, and another half of Cavity 2. A detailed view of a representative unit cell is presented in Figure 1(b), having inversion symmetry with respect to its center. The radius of the duct, denoted as $a$, is consistent for both the circular membranes and the cylindrical cavities. Cavity 1 and Cavity 2 have depths labeled as $\Delta_1$ and $\Delta_2$, respectively. The air within these cavities has density $\rho_a$ and sound speed $c_a$. The membranes are identical, clamped at their edges, and have uniform properties, namely, thickness $h_m$, material density $\rho_m$, and speed of elastic wave propagation $c_m$. The membrane thickness is considered negligible compared to the cavity depths ($h_A \ll \Delta_1, \Delta_2$), making the characteristic length of the unit cell equal to $\Delta_1 + \Delta_2$. We assume that the transverse vibration amplitudes of the linearly elastic membranes are small enough for linear infinitesimal elasticity to be applicable.

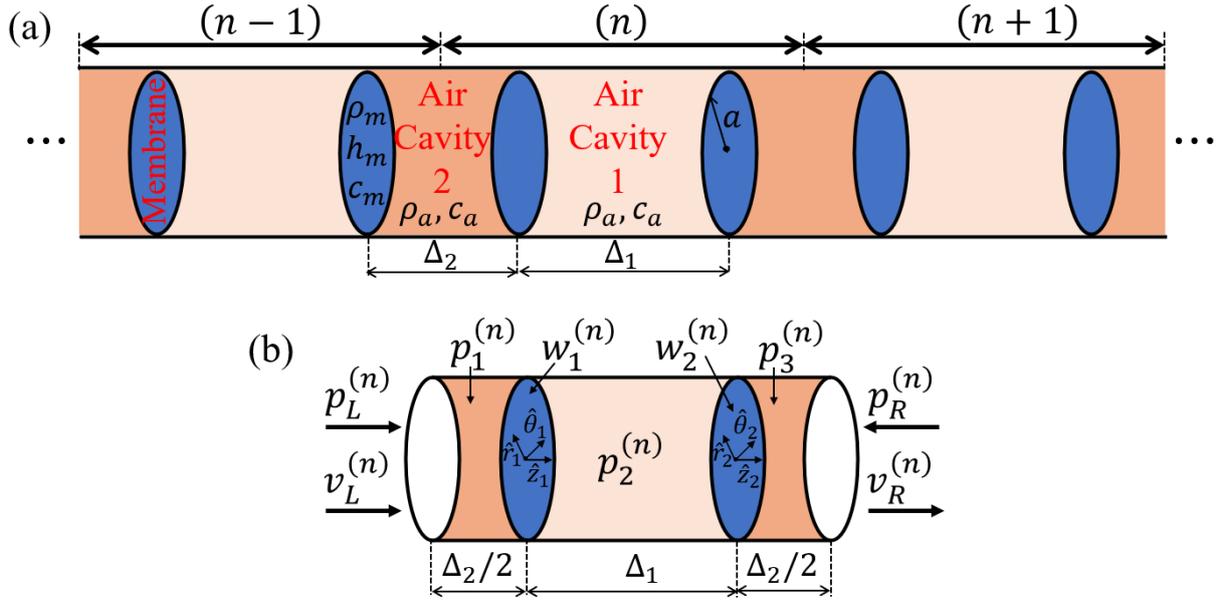

**Figure 1:** Schematic of the infinite vibroacoustic multilayered phononic metamaterial constructed by coupling alternating unit-cells (layers) consisting of half cavity 2 – membrane – cavity 1 – membrane – half cavity 2: (a) Three adjacent unit-cells, and (b) the $n^{th}$ unit-cell showing acoustic velocities and pressures at its left ($L$) and right ($R$) boundaries; blue circles represent the elastic linear membranes, while light or dark orange indicates the acoustic air cavity 1 or 2, respectively.

According to Bloch's theorem, the change in wave amplitude between adjacent unit cells is independent of their specific locations within the lattice. Therefore, a complete characterization of wave propagation in the infinite periodic assembly can be achieved solely through analyzing a representative unit cell. In Figure 1(b), the $n^{th}$ unit-cell is isolated from its neighboring cells, and its free body diagram is illustrated. The interaction between adjacent unit cells is described using the acoustic pressure and velocity pairs denoted by $p_L^{(n)}$ and $v_L^{(n)}$ at the left boundary ($L$), and $p_R^{(n)}$ and $p_R^{(n)}$ at the right boundary ($R$), respectively. In the following transfer matrix formulation, we assume that the acoustic pressure and velocity pair at the left boundary of the $n^{th}$ unit-cell is considered to be known, while the corresponding acoustic state at the right boundary is unknown.

A detailed analysis, based on the transfer matrix method, of a generalized version of this multilayered vibroacoustic lattice, where each unit cell is constructed from two distinct membranes and cavities, was recently reported in [48]. However, the selection of the unit cell was different from our choice in the current study. Here, the unit cell is shifted (compared to the case in [48]) to preserve inversion symmetry. As a result, while the band structure remains identical and independent of the unit cell selection, the acoustic pressure distribution along the unit cell, as well as the components of the transfer matrix in this study, will be different from those derived in [48]. Furthermore, altering the basis of the unit cell changes the geometric phase accumulated by the eigenmodes, such as changing the Zak phase from that of the unit cell in [48]. Therefore, for the inversion symmetry case selected here, the governing equations and boundary conditions will be detailed, followed by the final analytical solution for the coupled vibroacoustic phononic lattice.

## 2.2. Equations of motion and boundary conditions

Omitting dissipation effects, we denote by $p_1^{(n)}(r_1, \theta_1, z_1, t)$, $p_2^{(n)}(r_1, \theta_1, z_1, t)$ and $p_3^{(n)}(r_2, \theta_2, z_2, t)$ the instantaneous acoustic pressure distributions within the left, intermediate and right cavities, respectively, inside the $n^{th}$ unit-cell (see Figure 1(b)). These pressure distributions are governed by the following homogeneous 3D wave equations represented in cylindrical coordinates [51],

$$\nabla^2_{cyl(1)} p_1^{(n)} - \frac{1}{c_a^2}\frac{\partial^2 p_1^{(n)}}{\partial t^2} = 0, \qquad \nabla^2_{cyl(1)} p_2^{(n)} - \frac{1}{c_a^2}\frac{\partial^2 p_2^{(n)}}{\partial t^2} = 0$$
$$\nabla^2_{cyl(2)} p_3^{(n)} - \frac{1}{c_a^2}\frac{\partial^2 p_3^{(n)}}{\partial t^2} = 0 \qquad (1)$$

where $\nabla^2_{cyl(j)}$ is the Laplacian operator in the $j^{th}$ air cavity in cylindrical coordinates, i.e., $\nabla^2_{cyl(j)} = \frac{\partial^2}{\partial r_j^2} + \frac{1}{r_j}\frac{\partial}{\partial r_j} + \frac{1}{r_j^2}\frac{\partial^2}{\partial \theta_j^2} + \frac{\partial^2}{\partial z_j^2}$. Here we assume that the cylindrical duct is acoustically rigid in the radial direction, meaning that the radial component of the flow velocity at the duct's circumference is required to be zero. Additionally, the continuity of the normal acoustic velocity (along the $\hat{z}_{1,2}$ directions) at the cavity-membrane interfaces and the edges of the unit cell must be satisfied. Therefore, based on the linearized Euler transport equation, the following boundary conditions for the acoustic pressures inside the cavities are imposed:

$$\left.\frac{\partial p_1^{(n)}}{\partial r_1}\right|_{r_1=a} = \left.\frac{\partial p_2^{(n)}}{\partial r_1}\right|_{r_1=a} = \left.\frac{\partial p_3^{(n)}}{\partial r_2}\right|_{r_2=a} = 0$$
$$\left.p_1^{(n)}\right|_{z_1=-\frac{\Delta_2}{2}} = p_L^{(n)}, \quad -\frac{1}{\rho_a}\left[\int \frac{\partial p_1^{(n)}}{\partial z_1} dt\right]_{z_1=-\frac{\Delta_2}{2}} = v_{L_n}, \quad \left.\frac{\partial p_1^{(n)}}{\partial z_1}\right|_{z_1=0} = -\rho_a \frac{\partial^2 w_1^{(n)}}{\partial t^2}$$
$$\left.\frac{\partial p_2^{(n)}}{\partial z_1}\right|_{z_1=0} = -\rho_a \frac{\partial^2 w_1^{(n)}}{\partial t^2}, \quad \left.\frac{\partial p_2^{(n)}}{\partial z_1}\right|_{z_1=\Delta_1} = -\rho_a \frac{\partial^2 w_2^{(n)}}{\partial t^2} \qquad (2)$$
$$\left.\frac{\partial p_3^{(n)}}{\partial z_2}\right|_{z_2=0} = -\rho_a \frac{\partial^2 w_2^{(n)}}{\partial t^2}, \quad \left.p_3^{(n)}\right|_{z_2=\frac{\Delta_2}{2}} = -p_R^{(n)}, \quad -\frac{1}{\rho_a}\left[\int \frac{\partial p_3^{(n)}}{\partial z_2} dt\right]_{z_2=\frac{\Delta_2}{2}} = v_{R_n}$$

Here, $w_1^{(n)}(r_1, \theta_1, t)$ and $w_2^{(n)}(r_2, \theta_2, t)$ represent the instantaneous transverse displacements of the left and right membranes, respectively, inside the $n^{th}$ unit-cell. These forced transverse displacements of the membranes are governed by the following inhomogeneous 2D wave equations represented in polar coordinates,

$$\nabla^2_{pol(1)} w_1^{(n)} - \frac{1}{c_m^2}\frac{\partial^2 w_1^{(n)}}{\partial t^2} = \frac{1}{c_m^2 \rho_m h_m}\left(\left.p_2^{(n)}\right|_{z_1=0} - \left.p_1^{(n)}\right|_{z_1=0}\right)$$
$$\nabla^2_{pol(2)} w_2^{(n)} - \frac{1}{c_m^2}\frac{\partial^2 w_2^{(n)}}{\partial t^2} = \frac{1}{c_m^2 \rho_m h_m}\left(\left.p_3^{(n)}\right|_{z_2=0} - \left.p_2^{(n)}\right|_{z_1=\Delta_1}\right) \qquad (3)$$

where $\nabla^2_{pol(j)}$ is the Laplacian operator in the $j^{th}$ air cavity in polar coordinates, i.e., $\nabla^2_{pol(j)} = \frac{\partial^2}{\partial r_j^2} + \frac{1}{r_j}\frac{\partial}{\partial r_j} + \frac{1}{r_j^2}\frac{\partial^2}{\partial \theta_j^2}$. Moreover, we impose the following clamped boundary conditions for the membranes:

$$w_1^{(n)}\bigg|_{r_1=a} = w_2^{(n)}\bigg|_{r_2=a} = 0 \tag{4}$$

To analytically solve the coupled sound-structure problem defined by equations (1) to (4), we follow the approach in [49], where the transverse displacements of the membranes were explicitly solved without assuming eigenfunction (modal) expansions; as discussed in [49] this is important in order to satisfy the boundary conditions without resorting to infinite series expansions whose truncation is problematic for accuracy purposes. To this, the steady-state solution of the problem is explicitly derived for axisymmetric modes operating in the low-frequency range $0 < \Omega < \frac{\chi_1}{a} c_a$, where $\chi_1 = 3.8317$ is the leading root of the first order Bessel function of the first kind, $J_1$.

### 2.3. Analytical solution for the coupled sound-membrane vibroacoustic system

Let the steady-state axisymmetric acoustic pressure and velocity at the left edge of the $n^{th}$ unit-cell be expressed as:

$$p_{L_{(n)}} = P_{L_n} e^{i\Omega t}, \quad v_{L_{(n)}} = V_{L_n} e^{i\Omega t} \tag{5}$$

Where the amplitudes $P_{L_n}$ and $v_{L_n}$ are treated as inputs, and $\Omega$ is the constant operation frequency. Then, the corresponding explicit expressions for the acoustic pressure fields inside the air cavities of the unit cell are given by:

$$p_1^{(n)}(z_1, t) = \left( P_{1c} \cos\left(z_1 \frac{\Omega}{c_a}\right) + P_{1s} \sin\left(z_1 \frac{\Omega}{c_a}\right) \right) e^{i\Omega t}, \quad -\frac{\Delta_2}{2} \leq z_1 \leq 0$$

$$p_2^{(n)}(z_1, t) = \left( P_{2c} \cos\left(z_1 \frac{\Omega}{c_a}\right) + P_{2s} \sin\left(z_1 \frac{\Omega}{c_a}\right) \right) e^{i\Omega t}, \quad 0 \leq z_1 \leq \Delta_1 \tag{6}$$

$$p_3^{(n)}(z_2, t) = \left( P_{3c} \cos\left(z_2 \frac{\Omega}{c_a}\right) + P_{3s} \sin\left(z_2 \frac{\Omega}{c_a}\right) \right) e^{i\Omega t}, \quad 0 \leq z_2 \leq \frac{\Delta_2}{2}$$

In addition, the transverse deflection fields of the two membranes are given by:

$$w_1^{(n)}(r_1, t) = W_1 \left( \frac{J_0\left(\frac{a\Omega}{c_m}\right) - J_0\left(r_1 \frac{\Omega}{c_m}\right)}{J_2\left(\frac{a\Omega}{c_m}\right)} \right) e^{i\Omega t}$$

$$w_2^{(n)}(r_2, t) = W_2 \left( \frac{J_0\left(\frac{a\Omega}{c_m}\right) - J_0\left(r_2 \frac{\Omega}{c_m}\right)}{J_2\left(\frac{a\Omega}{c_m}\right)} \right) e^{i\Omega t} \tag{7}$$

Here, $J_0(\cdot)$ and $J_2(\cdot)$ are the Bessel functions of the first kind of order zero and two, respectively. The acoustic pressure and velocity at the right edge of the $n^{th}$ unit-cell (treated as outputs) are given by:

$$p_R^{(n)}(t) = P_{R_n} e^{i\Omega t}, \quad v_R^{(n)}(t) = V_{R_n} e^{i\Omega t} \tag{8}$$

Remarkably, expressions (6), (7), and (8) satisfy the equations of motion (1) and (3) as well as the boundary conditions (2) and (4). The explicit expressions for the pressure amplitudes $P_{jc}$ and $P_{jc}$ ($j = 1,2,3$), the membrane deflection amplitudes $W_1$ and $W_2$, and the amplitudes of the pressure and velocity at the right edge of the $n^{th}$ unit-cell $P_{R_n}$ and $V_{R_n}$, are listed in the Appendix. All these quantities explicitly depend on the acoustic pressure and velocity at the left edge of the $n^{th}$ unit-cell ($P_{L_n}$ and $V_{L_n}$), as well as the various parameters of the membranes and cavities composing the unit cell.

### 2.4. Transfer matrix formalism

To construct the *local transfer matrix* for a single unit cell, we denote the acoustic state vectors at the left and right boundaries of the $n^{th}$ unit-cell by $\mathbf{y}_{L(n)}$ and $\mathbf{y}_{R(n)}$, respectively, defined as:

$$\mathbf{y}_{L(n)} = \begin{bmatrix} p_{L(n)} \\ \rho_a c_a \\ v_{L(n)} \end{bmatrix}, \quad \mathbf{y}_{R(n)} = \begin{bmatrix} p_{R(n)} \\ \rho_a c_a \\ v_{R(n)} \end{bmatrix} \tag{9}$$

The acoustic velocity compatibility and the pressure equilibrium at the junction between the $n^{th}$ and the $(n+1)^{th}$ unit-cells suggest that:

$$\mathbf{y}_{L(n+1)} = \begin{bmatrix} p_{L(n+1)} \\ \rho_a c_a \\ v_{L(n+1)} \end{bmatrix} = \begin{bmatrix} -p_{R(n)} \\ \rho_a c_a \\ v_{R(n)} \end{bmatrix} \tag{10}$$

By definition, the local transfer matrix, denoted by $\mathbf{T}$, represents the recursion relation between the acoustic state vectors at the left boundaries of the $n^{th}$ and $(n+1)^{th}$ unit-cells as follows,

$$\mathbf{y}_{L(n+1)} = \mathbf{T}\mathbf{y}_{L(n)} \tag{11}$$

where the frequency-dependent components of $\mathbf{T}$, are explicitly defined as,

$$\mathbf{T} = \begin{bmatrix} T_{11} & T_{12} \\ T_{21} & T_{22} \end{bmatrix}$$

$$T_{11} = T_{22} = \cos(k_0(\Delta_1 + \Delta_2)) + \frac{\sigma_m}{\sigma_a}\sin(k_0(\Delta_1 + \Delta_2))$$
$$+ \frac{\sigma_m^2}{2\sigma_a^2}\sin(k_0\Delta_1)\sin(k_0\Delta_2)$$

$$T_{12} = i\left(-\sin(k_0(\Delta_1 + \Delta_2)) + 2\frac{\sigma_m}{\sigma_a}\cos\left(k_0\left(\Delta_1 + \frac{\Delta_2}{2}\right)\right)\cos\left(k_0\frac{\Delta_2}{2}\right)\right. \tag{12}$$
$$\left. + \frac{\sigma_m^2}{\sigma_a^2}\sin(k_0\Delta_1)\cos^2\left(k_0\frac{\Delta_2}{2}\right)\right)$$

$$T_{21} = -i\left(\sin(k_0(\Delta_1 + \Delta_2)) + 2\frac{\sigma_m}{\sigma_a}\sin\left(k_0\left(\Delta_1 + \frac{\Delta_2}{2}\right)\right)\sin\left(k_0\frac{\Delta_2}{2}\right)\right.$$
$$\left. + \frac{\sigma_m^2}{\sigma_a^2}\sin(k_0\Delta_1)\sin^2\left(k_0\frac{\Delta_2}{2}\right)\right)$$

where,

$$\sigma_m = \rho_m h_m \Omega^2 \frac{J_0\left(a\frac{\Omega}{c_m}\right)}{J_2\left(a\frac{\Omega}{c_m}\right)}, \quad k_0 = \frac{\Omega}{c_a}, \quad \sigma_a = \frac{\rho_a \Omega^2}{k_0} = \rho_a c_a \Omega.$$

As the considered system is time-invariant, linear, and possesses scalar material properties, the *reciprocity principle* holds [52], necessitating that the *local transfer matrix*, $\mathbf{T}$, satisfies that $det(\mathbf{T}) = T_{11}T_{22} - T_{12}T_{21} = 1$, i.e., $\mathbf{T}$ is a unimodular matrix regardless of the frequency $\Omega$. Moreover, the mirror symmetry property of the unit-cell with respect to its center yields that $T_{11} = T_{22}$.

## 2.5. Local impedance, scattering matrices, and reflection-transmission spectra

Rearranging the terms in equation (11) yields a relation between the acoustic pressures at the left and right boundaries of the $n^{th}$ unit cell and the acoustic velocities at the same locations. This relation is expressed by defining the *local impedance matrix*, denoted by $\mathbf{Z}$, in terms of the local transfer matrix components as follows:

$$\begin{bmatrix} \frac{p_{L_{(n)}}}{\rho_a c_a} \\ \frac{p_{R_{(n)}}}{\rho_a c_a} \end{bmatrix} = \mathbf{Z} \begin{bmatrix} v_{L_{(n)}} \\ v_{R_{(n)}} \end{bmatrix}, \quad \mathbf{Z} = \begin{bmatrix} Z_{11} & Z_{12} \\ Z_{21} & Z_{22} \end{bmatrix} = \begin{bmatrix} -\frac{T_{22}}{T_{21}} & \frac{1}{T_{21}} \\ \frac{1}{T_{21}} & -\frac{T_{11}}{T_{21}} \end{bmatrix} \qquad (13)$$

The symmetry of the local impedance matrix, $\mathbf{Z}$, directly reflects on the reciprocity principle.

Another way to describe the acoustical behavior of the vibroacoustic lattice is by using the *local scattering matrix*, denoted by $\mathbf{S}$. This matrix relates the incoming pressure wave vector $\left[p_{L_{(n)}}^+, p_{R_{(n)}}^-\right]^T$ and the outgoing pressure wave vector $\left[p_{L_{(n)}}^-, p_{R_{(n)}}^+\right]^T$ in terms of the components of the local transfer matrix, $\mathbf{T}$. Here, $p_{L_{(n)}}^+$ ($p_{L_{(n)}}^-$) and $p_{R_{(n)}}^+$ ($p_{R_{(n)}}^-$) represent the amplitudes of plane waves propagating in the positive (negative) $z$ direction at the left and right sides, respectively, of the $n^{th}$ unit-cell. A detailed derivation for a general expression of the local scattering matrix in terms of the local transfer matrix components and the scattering coefficients can be expressed as [47]:

$$\begin{bmatrix} p_{L_{(n)}}^- \\ p_{R_{(n)}}^+ \end{bmatrix} = \mathbf{S} \begin{bmatrix} p_{L_{(n)}}^+ \\ p_{R_{(n)}}^- \end{bmatrix}$$

$$\mathbf{S} = \begin{bmatrix} S_{11} & S_{12} \\ S_{21} & S_{22} \end{bmatrix} = \begin{bmatrix} \frac{T_{12} - T_{21}}{2T_{11} + T_{12} + T_{21}} & \frac{-2}{2T_{11} + T_{12} + T_{21}} \\ \frac{2}{2T_{11} + T_{12} + T_{21}} & \frac{T_{12} - T_{21}}{2T_{11} + T_{12} + T_{21}} \end{bmatrix} \qquad (14)$$

By construction, the local scattering matrix encodes the reflection and transmission coefficients, respectively denoted by $r$ and $t$, for the inversion symmetric case as follows:

$$r(\Omega) = |S_{11}|^2 = |S_{22}|^2 = \left|\frac{T_{12} - T_{21}}{2T_{11} + T_{12} + T_{21}}\right|^2$$

$$t(\Omega) = |S_{12}|^2 = |S_{21}|^2 = \frac{4}{|2T_{11} + T_{12} + T_{21}|^2} \qquad (15)$$

## 2.6. Band structure characterization

The band structure of the vibroacoustic phononic metamaterial with infinite unit cells is fully characterized using the Bloch-Floquet theorem, where the eigenvalues of the local transfer matrix are given by:

$$\Lambda_{1,2} = e^{\pm i\mu(\Omega)} \qquad (16)$$

where $\mu$ denotes the (frequency dependent) *propagation constant*, given by:

$$\mu(\Omega) = arccos\left(\frac{1}{2}tr(\mathbf{T})\right) = arccos(T_{11}) \qquad (17)$$

Moreover, the corresponding eigenvectors of matrix $\mathbf{T}$, denoted by $X_1$ and $X_2$, are given by:

$$X_{1,2} = \begin{bmatrix} T_{12} \\ e^{\pm i\mu} - T_{11} \end{bmatrix} = \begin{bmatrix} T_{12} \\ \pm i\sqrt{1 - T_{11}^2} \end{bmatrix} \quad (18)$$

According to equation (16), wave propagation occurs when $\mu$ is strictly real, corresponding to $-1 \leq T_{11} \leq 1$, which defines the *passbands* where the eigenvalues of the transfer matrix satisfy $|\Lambda_{1,2}(\Omega)| = 1$. In contrast, wave attenuation arises when $\mu = i\theta$ or $\mu = \pi + i\theta$, with $T_{11} < -1$ or $T_{11} > 1$ defining the *stopbands*, where one eigenvalue satisfies $|\Lambda(\Omega)| > 1$ and the other $|\Lambda(\Omega)| < 1$. The cut-on and cut-off frequencies, obtained by requiring $T_{11} = \pm 1$, are determined by solving the following transcendental characteristic equation:

$$cos(k_0(\Delta_1 + \Delta_2)) + \frac{\sigma_m}{\sigma_a} sin(k_0(\Delta_1 + \Delta_2)) + \frac{\sigma_m^2}{2\sigma_a^2} sin(k_0\Delta_1)sin(k_0\Delta_2) = \pm 1 \quad (19)$$

## 3. Vibroacoustic topological interface states – The Lattice of infinite extent

In this section, we outline a systematic approach to constructing vibroacoustic topological interface states by examining topological phase transitions through band structure evolution and symmetry analysis of band-edge pressure distributions. We then analyze the interface states in infinite lattices using the Zak phase, surface impedance, and reflection coefficients. Rather than relying on the straightforward expression of the Zak phase, we approach it through the transfer matrix perspective. While the Zak phase is a well-established tool for characterizing band topology, its abstract nature complicates the design of interface modes, particularly in complex systems like the proposed vibroacoustic system. To make the design of topological states more practical and physically insightful, we adopt the surface impedance concept. Through the transfer matrix formalism, we establish explicit criteria for the existence of topological interface states and, remarkably, derive a direct relation for locating these states based on the system's physical parameters. Our analysis also uncovers new insights into the role of reflection coefficients in determining the location of interface states.

### 3.1. Construction of topological interface states

To study the topological properties of the multilayered vibroacoustic phononic metamaterial, we introduce a contrast parameter $\gamma$ such that the depth of each of the two cavities of the unit-cell is expressed as,

$$\begin{aligned} \Delta_1 &= \Delta_0(1 + \gamma) \\ \Delta_2 &= \Delta_0(1 - \gamma) \end{aligned} \quad (20)$$

where $\Delta_0$ is a nominal cavity depth, and $\gamma$ signifies the asymmetry in the cavity depths. Importantly, by just changing the contrast parameter, $\gamma$, the length of the unit-cell remains invariant and equal to $2\Delta_0$.

The influence of the contrast parameter, $\gamma$, on the bounding frequencies of the six leading bands/bandgaps is illustrated in Figure 2 for nominal cavity depth $\Delta_0 = 6[cm]$. The other system parameters are listed below in Table I. The bounding frequencies are determined as the roots of equation (19). In the considered sub-wavelength scale, the resulting bandgaps are categorized into three types, namely, *plasma*, local resonance, and Bragg-like band-splitting induced bandgaps. The latter, recently studied in [48], was found to be a generalization of band-folding induced bandgaps. For a detailed explanation of each bandgap type and the underlying physical mechanisms generating them, the reader is referred to [48]. Remarkably, all the bandgaps observed in Figure 2 are located well below the first-order Bragg diffraction frequency, given by $\Omega_{Bragg}^{(1)} = \frac{c_a}{(\Delta_1+\Delta_2)}$ [$Hz$], classifying them as *sub-wavelength bandgaps*. In physical terms, the membranes act as local resonators, with the local resonance bandgaps forming in small neighborhoods of their natural frequencies. Based on equation (7), the $n^{th}$ natural frequency of the membrane, denoted by $\Omega_m^{(n)}$, can be expressed as,

$$\Omega_m^{(n)} = \frac{c_m}{a}\beta_2^{(n)} \qquad (21)$$

where, $\beta_2^{(n)}$ is the $n^{th}$ root of the second-order Bessel function of the first kind. This relation yields an infinite count of natural frequencies for the in-air membranes.

In addition to preserving the unit-cell length, i.e., $\Delta_1 + \Delta_2 = 2\Delta_0$, the mirror symmetry of the unit-cell is also preserved as $\gamma$ varies within the range $-1 < \gamma < 1$. The plot in Figure 2 indicates that the bounding frequencies of (theoretically infinite in number) bands can be adjusted to intersect simultaneously, with symmetry-enforced degeneracy at $\gamma = 0$. The intersections (i.e., band-crossings) occur between the $(2n-1)^{th}$ and $(2n)^{th}$ bands, with $n = 1,2,\cdots$. At the degeneracy point, $\gamma = 0$, the unit cell effectively consists of two identical membrane-cavity layers, leading to Brillouin zone folding.

Consistently, the band-crossing phenomenon in Figure 2 occurs exclusively within the Bragg-like band-splitting induced bandgaps (or within the classical Bragg bandgaps that appear at high-frequencies) and not within the local resonance bandgaps. Within the range $-1 < \gamma < 1$, the closing and reopening of band-splitting induced bandgaps, accompanied by the crossing of the bands, is analogous to the *band inversion* process observed in electronic and photonic systems. Thus, *the local resonance bandgaps are expected to remain topologically identical, while the band-splitting induced bandgaps could undergo topological phase transitions*, with $\gamma = 0$ marking a topological transition point in the multilayered vibroacoustic system. Consequently, phononic crystals (PCs) with different values of $\gamma$ on either side of the transition point exhibit distinctly different topological characteristics within their respective band-splitting induced bandgaps. Without loss of generality, from here on unit-cells with $\gamma = 1/3$ ($\Delta_1 = 8[cm], \Delta_2 = 4[cm]$) and $\gamma = -1/3$ ($\Delta_1 = 4[cm], \Delta_2 = 8[cm]$), will be denoted as phononic crystal type I (PC-I) and phononic crystal type II (PC-II), respectively. These are selected to represent two topologically distinct vibroacoustic phononic crystals, as indicated by the vertical dashed-black lines in Figure 2. It is evident that both PC-I and PC-II unit-cells share the same bounding frequencies, resulting in identical band structures (see Figure 3). This is mathematically supported by the fact that they have the same propagation constant, $\mu$, as given in equations (12) and (17). However, infinite vibroacoustic lattices composed uniformly of PC-I or PC-II unit-cells exhibit distinct eigenvectors, as expressed in equations (12) and (18), which encode their different topological properties, as will be demonstrated subsequently.

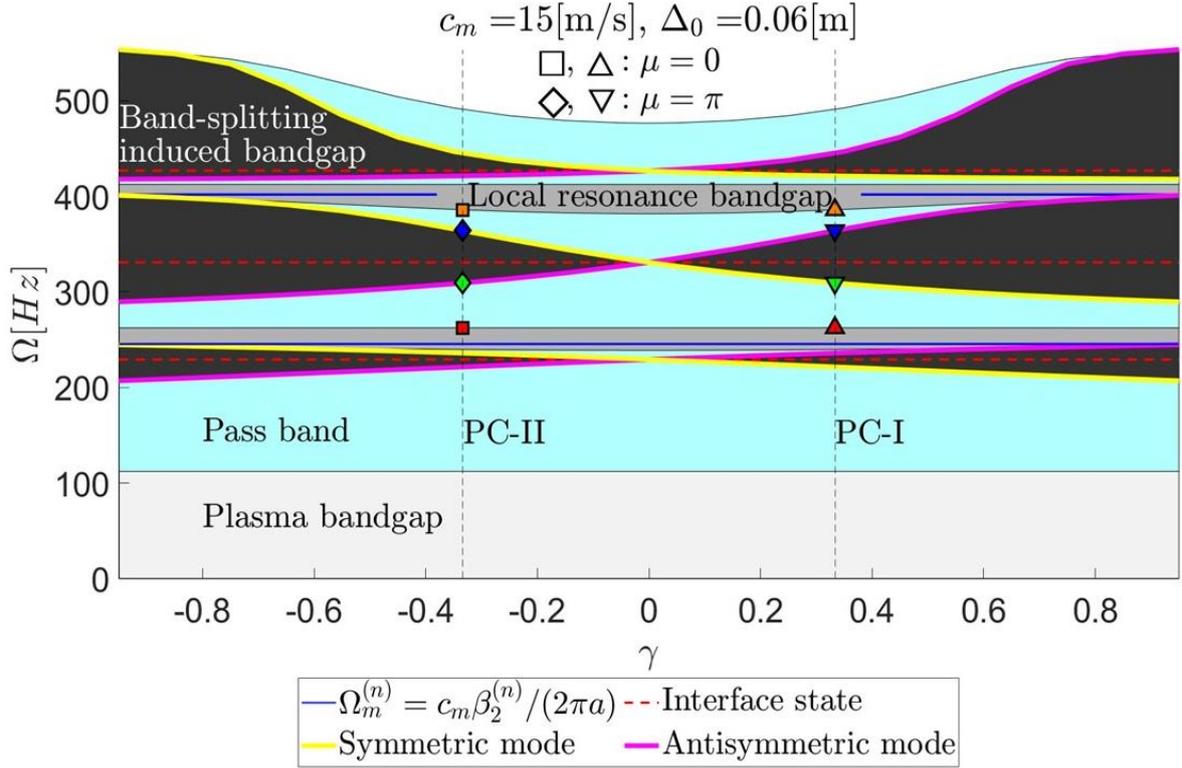

**Figure 2**: The six leading bandgaps/bands in the multilayered vibroacoustic metamaterial for varying contrast parameter ($\gamma$), showing the topological phase transition (see text for parameters). Cyan strips indicate pass bands, while the three levels of gray-scale shading, from light to dark, indicate the plasma, local resonance, and band-splitting induced bandgaps, respectively. Solid blue lines represent the natural frequencies of the membrane, dashed red lines the interface state (discussed later), and solid yellow or magenta lines, the symmetric (even) or antisymmetric (odd) eigenmodes, respectively. Black dashed columns denote two topologically different phononic crystals, denoted by PC-I and PC-II. Red upward-pointing triangle (square), green downward-pointing triangle (diamond), blue downward-pointing triangle (diamond), and orange upward-pointing triangle (square) indicate the eigenfrequencies (bounding frequencies) for the PC-I or PC-II configurations, respectively, see Figs. 3 and 4.

Prior to examining the topological structures of infinite vibroacoustic lattices composed of PC-I or PC-II unit-cells, we utilize the derived analytical pressure distribution inside the three cavities along a representative unit-cell, as provided in equation (6), to assess the symmetry of the eigenmodes at the bounding frequencies of the band-splitting induced bandgaps (i.e., band-edge states). To this end, the eigenmodes corresponding to $\mu = \pi$ and exhibiting the band inversion phenomenon, can be classified into *symmetric* or *antisymmetric* modes, as indicated by the yellow and magenta curves, respectively, in Figure 2. This classification procedure is based on the pressure profile inside a representative unit-cell, as will be shown in Figure 4. The evolution of the interface states as $\gamma$ varies, depicted by the dashed red line in Figure 2, will be analyzed and discussed subsequently.

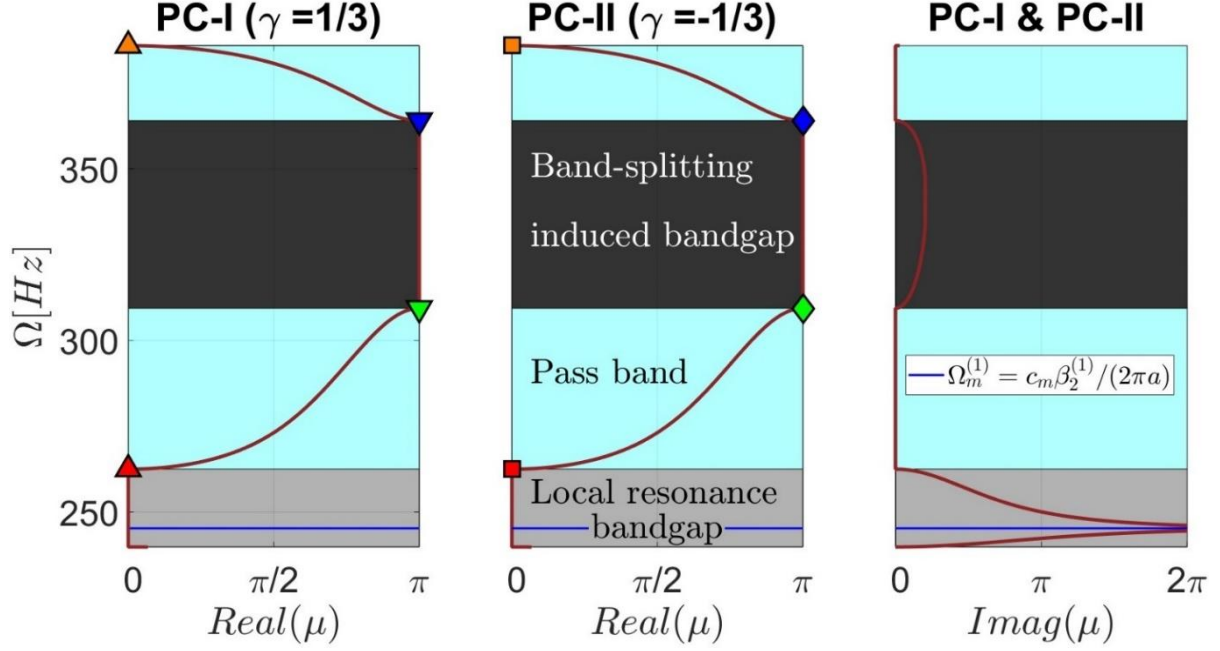

**Figure 3**: The third and fourth bands/bandgaps of the band structure for infinite lattices composed of uniform PC-I or PC-II unit-cells. Real and imaginary parts of the propagation constant, $\mu$, shown in dark brown, are calculated based on equation (17). The system parameters, marked bounding frequencies, and color interpretations are consistent with Fig. 2.

To visualize the acoustic pressure eigenfunctions (mode shapes) within a representative unit-cell for PC-I and PC-II lattice configurations, we selected the four bounding frequencies that determine the edges of the third and fourth bands, i.e., band-edge states. These band-edge states are marked in Figure 2 by upward- and downward-pointing triangles for the PC-I configuration, and by squares and diamonds for the PC-II configuration. The exact locations of these selected band-edge states projected on the dispersion curves are illustrated in Figure 3 for both PC-I and PC-II lattice configurations, where the third and fourth bands/bandgaps of the band structure are presented. Visualizations of the corresponding pressure mode shapes, within a representative unit-cell for the PC-I and PC-II configurations, at these selected band-edge states frequencies are illustrated in Figure 4. For the PC-I configuration, the pressure eigenfunction associated with the cut-off frequency of the third band (green downward-pointing triangle) is symmetric, while the pressure eigenfunction associated with the cut-on frequency of the fourth band (blue downward-pointing triangle) is antisymmetric. However, these characteristics are inverted for the PC-II configuration, where the frequency of the symmetric mode becomes higher than that of the antisymmetric one. *Therefore, PC-I and PC-II lattice configurations are topologically distinct despite their apparently identical band structures*. The observed switching of the band-edge states is attributed to the band-inversion phenomenon, which has been previously exploited in electronic [3], photonic [44], and phononic [29] crystals to achieve localized states. Such localized states appear at the interface of two topologically distinct lattices, in the present case one with $\gamma > 0$ (PC-I) and the other with $\gamma < 0$ (PC-II). While the band-inversion phenomenon is evident across the band-edge states corresponding to $\mu = \pi$, it is absent at the band-edge states for $\mu = 0$, as presented in Figure 3. These states are topologically identical for both PC-I and PC-II lattice configurations, as can be seen from the pressure distribution depicted in Figure 4.

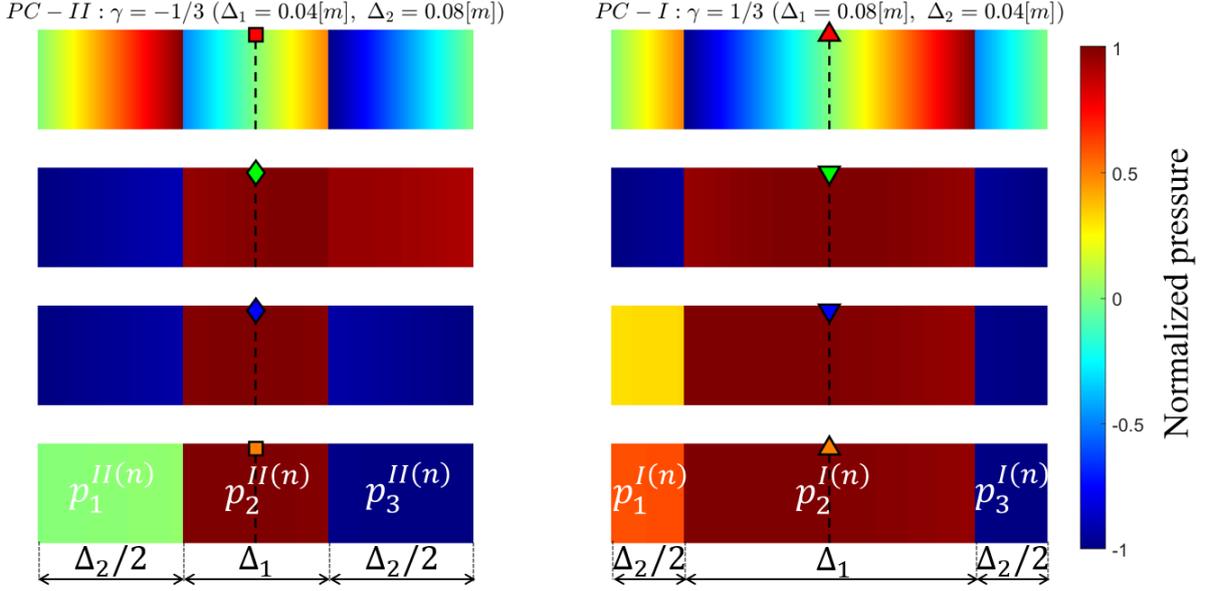

**Figure 4**: Acoustic pressure eigenfunctions (mode shapes) inside the three cavities along a representative unit cell, computed at the four bounding frequencies marked in Figs. 2 and 3 for the PC-I lattice configuration (right column) and the PC-II lattice configuration (left column).

Based on the aforementioned discussion, topological interface states can be realized at the interface of two connected lattices with different topological phases within a common bandgap. Figure 5 depicts the schematic of the composite lattice with the interface formed by two semi-infinite lattices with topologically distinct unit-cell configurations; namely, a right lattice composed of PC-I unit-cells ($\gamma > 0$) and a left lattice composed of PC-II unit-cells ($\gamma < 0$). A convenient way to distinguish between the two PC-I and PC-II semi-infinite lattice configurations is by simply exchanging the cavity depths, i.e., inverting $\Delta_1 \leftrightarrow \Delta_2$.

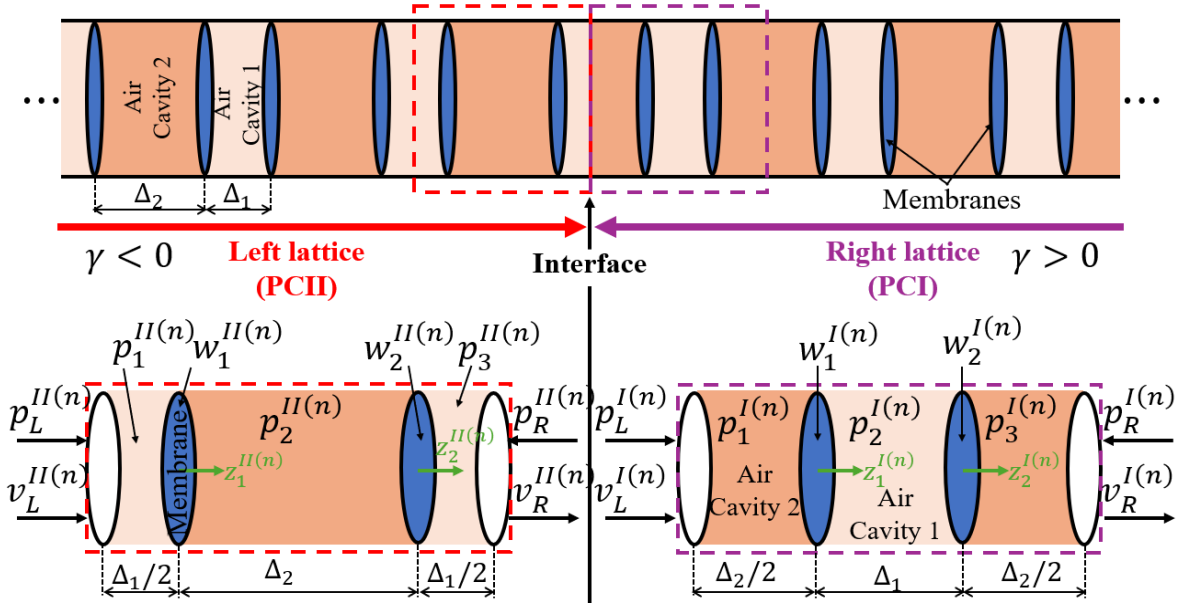

**Figure 5**: Schematic of the composite vibroacoustic lattice, composed of two semi-lattices with PC-I (right) and PC-II (left) unit-cells.

Superscripts "I" and "II" are used in Figure 5 to distinguish between the right lattice of PC-I unit-cells, and the left one of PC-II unit-cells, respectively. In the subsequent analysis, we will explore the physical mechanisms behind the formation of topological interface states in this composite lattice system from two perspectives: First, by examining the properties of the infinite lattice, and second, by analyzing the finite lattice formed by a limited number of PC-I and PC-II unit-cells.

### 3.2. Vibroacoustic topological interface states

In this section we employ the previous transfer matrix derivations to investigate the topological interface states in the composite lattice of infinite extent. To this end, let $T^I$ and $T^{II}$ denote the local transfer matrices for the infinite lattices composed of PC-I and PC-II unit-cells, respectively. These unimodular matrices can be expressed, based on (12), as:

$$T^I = \begin{bmatrix} T^I_{11} & T^I_{12} \\ T^I_{21} & T^I_{22} \end{bmatrix}, \qquad T^{II} = \begin{bmatrix} T^{II}_{11} & T^{II}_{12} \\ T^{II}_{21} & T^{II}_{22} \end{bmatrix}$$

$$T^I_{11} = T^I_{22} = T^{II}_{11} = T^{II}_{22} =$$
$$= cos(k_0(\Delta_1 + \Delta_2)) + \frac{\sigma_m}{\sigma_a} sin(k_0(\Delta_1 + \Delta_2))$$
$$+ \frac{\sigma_m^2}{2\sigma_a^2} sin(k_0\Delta_1) sin(k_0\Delta_2)$$

$$T^I_{12} = i\left(-sin(k_0(\Delta_1 + \Delta_2)) + \frac{2\sigma_m}{\sigma_a} cos\left(k_0\left(\Delta_1 + \frac{\Delta_2}{2}\right)\right) cos\left(k_0 \frac{\Delta_2}{2}\right)\right.$$
$$\left. + \frac{\sigma_m^2}{\sigma_a^2} sin(k_0\Delta_1) cos^2\left(k_0 \frac{\Delta_2}{2}\right)\right)$$

$$T^{II}_{12} = i\left(-sin(k_0(\Delta_1 + \Delta_2)) + \frac{2\sigma_m}{\sigma_a} cos\left(k_0\left(\Delta_2 + \frac{\Delta_1}{2}\right)\right) cos\left(k_0 \frac{\Delta_1}{2}\right)\right. \quad (22)$$
$$\left. + \frac{\sigma_m^2}{\sigma_a^2} sin(k_0\Delta_2) cos^2\left(k_0 \frac{\Delta_1}{2}\right)\right)$$

$$T^I_{21} = -i\left(sin(k_0(\Delta_1 + \Delta_2)) + 2\frac{\sigma_m}{\sigma_a} sin\left(k_0\left(\Delta_1 + \frac{\Delta_2}{2}\right)\right) sin\left(k_0 \frac{\Delta_2}{2}\right)\right.$$
$$\left. + \frac{\sigma_m^2}{\sigma_a^2} sin(k_0\Delta_1) sin^2\left(k_0 \frac{\Delta_2}{2}\right)\right)$$

$$T^{II}_{21} = -i\left(sin(k_0(\Delta_1 + \Delta_2)) + 2\frac{\sigma_m}{\sigma_a} sin\left(k_0\left(\Delta_2 + \frac{\Delta_1}{2}\right)\right) sin\left(k_0 \frac{\Delta_1}{2}\right)\right.$$
$$\left. + \frac{\sigma_m^2}{\sigma_a^2} sin(k_0\Delta_2) sin^2\left(k_0 \frac{\Delta_1}{2}\right)\right)$$

In the following analysis we aim to study the emergence of localized topological modes at the interface of the two semi-infinite lattices (see Fig. 5) employing different methodologies. We start with the classical approach based on the Zak phase.

### 3.2.1. Zak phase computation

The geometric phase concept, specifically the *Zak phase* [41], is recognized as a well-established method for understanding topological phase transitions and classifying the topological properties of bulk bands in 1D periodic systems. The Zak phase quantifies the geometric phase acquired over the Brillouin zone, enabling the classification of bulk bands in phononic systems. This quantization is directly tied to the emergence of topologically protected interface states, that is, of localized modes that appear at the boundaries between regions with distinct topological characteristics. Specifically, the Zak phase for the $n^{th}$ band, denoted by $\theta_n^{Zak}$, is defined as:

$$\theta_n^{Zak} = \int_{-\pi}^{\pi} \left[ i \int_{Unit\ cell} \frac{1}{2\rho_a c_a^2} dr dz u_{n,\mu}^*(r,z) \partial_\mu u_{n,\mu}(r,z) \right] d\mu \qquad (23)$$

where $u_{n,\mu}(r,z)$ is the normalized Bloch pressure characteristic function when the wavenumber of the $n^{th}$ frequency band in the periodic unit is $\mu$, $u_{n,\mu}^*(r,z)$ is the conjugate transpose of the characteristic function $u_{n,\mu}(r,z)$, and the factor $(1/2\rho_a c_a^2)$ is the weight function of the acoustic system and is a real constant. Physically, Zak phase encodes information about the global structure of the wavefunction and the underlying symmetry of the system. In general, the Zak phase can assume any value when the unit cell is chosen arbitrarily [41,53]. However, when the unit cell exhibits mirror symmetry (as in the considered case), the Zak phase is restricted to quantized values of either 0 or $\pi$ [53].

Due to the inversion symmetry of the unit-cell and the derived analytical solution for the sound-membrane interaction, the Zak phase for the $n^{th}$ band can be determined without solving the integral in equation (23), by employing two alternative approaches. The first approach involves assessing the symmetry of the pressure eigenfunctions at the band-edge states, specifically at $\mu = 0$ and $\mu = \pi$ [44]. If the band-edge states of a given band exhibit different symmetries, the Zak phase of that band is $\pi$. Conversely, if the band-edge states share the same symmetry, the Zak phase is 0. For example, the third band of PC-I has band-edge states with different symmetries (see Figures 3 and 4), indicating a Zak phase of $\pi$. In contrast, the fourth band of PC-I has band-edge states with the same antisymmetric behavior, resulting in a Zak phase of 0.

The second approach for determining the Zak phase of the inversion-symmetric unit-cell relies on the components of the eigenvectors of the local transfer matrices, $X_{1,2} = [T_{12} \quad e^{\pm i\mu} - T_{11}]^T$ in relations (18). Indeed, as demonstrated in [44], if $T_{12}$ and $e^{\pm i\mu} - T_{11}$ simultaneously become zero at the frequency $\widetilde{\Omega}_n$ within the $n^{th}$ band, then the Zak phase of the $n^{th}$ band is $\pi$. Otherwise, if $T_{12}$ and $e^{\pm i\mu} - T_{11}$ do not become simultaneously zero within the $n^{th}$ band, the Zak phase is 0.

Considering that $T_{11} = T_{22} = cos(\mu)$ and $det(\mathbf{T}) = 1$, the determination of the Zak phase for the $n^{th}$ band based on the components of the eigenvectors simplifies to,

$$\theta_n^{Zak} = \begin{cases} \pi, & if\ T_{12}(\widetilde{\Omega}_n) = 0 \\ 0, & otherwise \end{cases} \qquad (24)$$

where $\widetilde{\Omega}_n$ is a bounding frequency of the $n^{th}$ band, i.e., $\widetilde{\Omega}_n = \Omega_n^{cut-on}$ or $\Omega_n^{cut-off}$ satisfying $T_{11}(\widetilde{\Omega}_n) = \pm 1$. Equation (24) applies to any 1D phononic unit-cell with inversion symmetry and demonstrates the efficacy of the transfer matrix method in determining the Zak phase. Moreover, it is interesting to note that, while the band structure is fully encoded in $tr(\mathbf{T}) = T_{11} + T_{22}$, the topological characterization of the bands is directly related to the components $T_{12}$ and $T_{21}$. Specifically, for the unit-cells with PC-I and PC-II, it can be observed from equation (22) that $T_{12}^I \neq T_{12}^{II}$ and $T_{21}^I \neq T_{21}^{II}$, respectively. Consequently, their Zak phases across the bands are not identical, indicating that they are topologically distinct. To the best of our knowledge, such straightforward recipe (24) for determining the Zak phase in 1D unit-cells (phononic crystals) satisfying inversion-symmetry, based solely on the local transfer matrix component $T_{12}$ at the band-edge states, has not been previously reported.

After determining the Zak phase for each band, the existence of an interface state is established by the summation of Zak phases below the gap. Specifically, if two unit-cells with different sums of Zak phases below the $n^{th}$ bandgap are coupled, an edge state localized at the interface within the $n^{th}$ bandgap exists [44]. Hence, the existence of a topological interface state at the interface of the composite vibroacoustic lattice of infinite extent of Figure 5 is established.

### 3.2.2. Surface impedance

To bring the design of topological states to a practical level, providing physical insights and predictive design tools, we now consider an alternative approach for establishing topological interface modes in the infinite composite lattice based on surface impedance. The concept of surface impedance, which encodes surface scattering properties, was introduced in [44] to describe the topological properties of bulk dispersion through the geometric (Zak) phases of the bulk bands. Remarkably, a rigorous relation was derived that relates the surface impedance of a 1D PC to its bulk properties through the geometric (Zak) phases of the bulk bands. This relation can be employed to determine the existence or absence of interface states at the interface between two PCs within a specific bandgap, as well as to determine the exact frequency of any such states if they exist. Specifically, it was shown that the condition for the formation of an interface state in the interfacial region between two 1D PCs, i.e., PC-I and PC-II in this case, is simply [44],

$$Z_{ij}^{I} + Z_{ij}^{II} = 0 \tag{25}$$

where $Z_{ij}^{I}$ ($Z_{ij}^{II}$) is the $ij$-th component of the local impedance matrix (surface impedance) of the semi-infinite lattice composed of PC-I (PC-II) unit-cells, located on the right-hand (left-hand) side of the interface. Condition (25) implies that, *to guarantee the existence of a topological interface state, the surface impedances in the left and right semi-infinite lattices must have opposite signs within a common bandgap*. It is worth noting that the signs of the surface impedance components $Z_{ij}^{I}$ and $Z_{ij}^{II}$, $i,j = 1,2$, remain unchanged within a bandgap according to the following proposition.

***Proposition***: For the composite vibroacoustic lattice of infinite extent depicted in Figure 5, within an arbitrary $n^{th}$ bandgap, the signs of the surface impedance components $Z_{ij}^{I}$ and $Z_{ij}^{II}$, $i,j = 1,2$, remain unchanged.

***Proof***: According to equation (13), it is evident that the surface impedance components, $Z_{11}$, $Z_{12}$, $Z_{21}$ and $Z_{22}$, possess constant signs as long as the signs of the local transfer matrix components $T_{11}$, $T_{12}$, $T_{21}$ and $T_{22}$, remain unchanged. Therefore, to prove this proposition, we need to show that the signs of the corresponding local transfer matrix components do not change within an arbitrary $n^{th}$ bandgap. The inversion symmetry property of the unit-cell provides that $T_{11} = T_{22}$. Therefore, from equation (17), it is evident that within a bandgap, $T_{11} < -1$ or $T_{11} > 1$. Hence, both $T_{11}$ and $T_{22}$ do not cross zero inside the bandgap, meaning their signs remain unchanged. The same can be demonstrated for the components $T_{12}$ and $T_{21}$ when considering the fact that within the bandgap it holds that $T_{11}^2 > 1$. Then, considering the reciprocity principle yielding $det(\boldsymbol{T}) = 1$, which holds at any frequency, including those within the bandgap, we obtain $T_{11}^2 - 1 = T_{12}T_{21} > 0$. This implies that both $T_{12}$ and $T_{21}$ are either both positive or both negative at any frequency within the bandgap. Therefore, no change of sign can occur. ∎

To determine the surface impedance components $Z_{ij}^I$ and $Z_{ij}^{II}$, we introduce the corresponding local transfer matrices $\boldsymbol{T}^I$ and $\boldsymbol{T}^{II}$ from equation (22) into the expressions for the local impedance matrix given in equation (13). Instead of using the surface impedances, the criterion for the existence of an interface state at the interfacial region within the $n^{th}$ bandgap can be expressed solely in terms of the components $T_{21}^I$ and $T_{21}^{II}$, of the local transfer matrices as follows,

$$\begin{cases} \text{Existence of interface state:} & if\ sgn\left(\Im(T_{21}^I(\Omega_n))\right) \neq sgn\left(\Im(T_{21}^{II}(\Omega_n))\right) \\ \text{Nonexistence of interface state:} & if\ sgn\left(\Im(T_{21}^I(\Omega_n))\right) = sgn\left(\Im(T_{21}^{II}(\Omega_n))\right) \end{cases} \quad (26)$$

where "Existence" ("Nonexistence") indicates the presence (absence) of an interface state within the $n^{th}$ bandgap. Additionally, $\Im(\psi)$ represents the imaginary part of $\psi$, and $\Omega_n$ is a frequency within the $n^{th}$ bandgap, i.e., $\Omega_n \in \left(\Omega_{gap,n}^{cut-on}, \Omega_{gap,n}^{cut-off}\right)$. Condition (26) indicates that for an interface state to exist in the interface of the composite vibroacoustic lattice of Figure 5, the corresponding local transfer matrix components $T_{21}^I$ and $T_{21}^{II}$ must have opposing signs within the same (the $n^{th}$) bandgap.

To determine the location (i.e., the frequency) of the interface state, denoted by $\Omega_n^*$, within the $n^{th}$ bandgap, we consider the surface impedance components $Z_{ij}^I$ and $Z_{ij}^{II}$ in terms of the local transfer matrices into equation (25). It can be shown that the following relation holds:

$$T_{21}^I(\Omega_n^*) + T_{21}^{II}(\Omega_n^*) = 0 \quad (27)$$

When an interface state exists within the $n^{th}$ bandgap (based on criterion (26)), relation (27) determines the precise location of the interface state in terms of the local transfer matrices of the two coupled vibroacoustic semi-infinite lattices composed of PC-I and PC-II unit-cells. Considering the inversion symmetry of the unit-cells, i.e., $T_{11}^I = T_{22}^I = T_{11}^{II} = T_{22}^{II}$, as well as the unimodularity of the local transfer matrices, i.e., $det(\boldsymbol{T}^I) = det(\boldsymbol{T}^{II}) = 1$, one obtains the following relation:

$$T_{12}^I(\Omega)T_{21}^I(\Omega) = T_{12}^{II}(\Omega)T_{21}^{II}(\Omega), \quad \forall \Omega \quad (28)$$

Relation (28) holds for any 1D unit-cell with inversion symmetry, regardless of the interface states. The combination of equations (27) and (28) provides the following relations that *must be satisfied at an interface state in the composite lattice of infinite extent with inversion-symmetric unit cells*:

$$\begin{array}{l} T_{21}^I(\Omega_n^*) + T_{21}^{II}(\Omega_n^*) = 0 \\ T_{12}^I(\Omega_n^*) + T_{12}^{II}(\Omega_n^*) = 0 \end{array} \quad (29)$$

To the best of our knowledge, such straightforward relations characterizing the interface states in terms of the local transfer matrix components have not been previously reported. A physical interpretation of relations (29) will be provided in the next subsection.

After introducing the components of the local transfer matrices from equations (22) to (29) and performing algebraic manipulations, the following transcendental frequency equation for determining the exact location of the interface state within the $n^{th}$ bandgap ($\Omega_n^*$) is obtained,

$$tan\left(\frac{\Omega_n^*}{c_a}\Delta_0\right) + \frac{2\rho_a c_a J_2\left(\frac{\Omega_n^*}{c_m}a\right)}{\rho_m h_m \Omega_n^* J_0\left(\frac{\Omega_n^*}{c_m}a\right)} = 0 \qquad (30)$$

where, the relations for $\sigma_m$, $\sigma_a$ and $k_0$ are taken from equation (12). Additionally, $\Delta_1$ and $\Delta_2$, are expressed as functions of the contrast parameter $\gamma$ and the nominal cavity depth $\Delta_0$ from equation (20). Equation (30) reveals that *the frequency of an interface state is independent of the contrast parameter $\gamma$*. This observation is consistent with the results shown in Figure 2, where the horizontal dashed red lines represent the interface states with no dependence on $\gamma$. Remarkably, equation (30) offers a direct tool for the predictive design and adjustment of interface states based on the system's physical parameters. To the best of our knowledge, such a direct relationship has not been previously reported for either realistic or even discrete systems. This predictive design tool is expected to enable a deeper understanding of sound-matter interactions at a fundamental level and to harness the power of topology for energy management in vibroacoustic systems, paving the way for a new generation of lightweight and controllable sound-based structures. Additionally, the procedure proposed here to derive equation (30) using the transfer matrix formalism is not limited to vibroacoustic systems; it can be applied to any photonic or phononic system with an inversion-symmetric unit cell. This broad applicability emphasizes the significance of our methodology, making it a powerful and versatile tool for analyzing and designing topological states across various physical systems.

### 3.2.3. Reflection coefficients

Lastly, further perspective on the physical mechanism governing the formation of interface states in the composite vibroacoustic lattice of Figure 5, can be gained from the viewpoint of reflection coefficients for the component right (I) and left (II) semi-infinite lattices. From equation (15), the corresponding reflection coefficients, denoted by $r^I$ and $r^{II}$, respectively, are expressed as follows:

$$r^I(\Omega) = \left|\frac{T_{12}^I(\Omega) - T_{21}^I(\Omega)}{2T_{11}^I(\Omega) + T_{12}^I(\Omega) + T_{21}^I(\Omega)}\right|^2$$
$$r^{II}(\Omega) = \left|\frac{T_{12}^{II}(\Omega) - T_{21}^{II}(\Omega)}{2T_{11}^{II}(\Omega) + T_{12}^{II}(\Omega) + T_{21}^{II}(\Omega)}\right|^2 \qquad (31)$$

In addition, the interface state relations given in (29) can be rearranged as follows:

$$T_{12}^I(\Omega_n^*) - T_{21}^I(\Omega_n^*) = -\left(T_{12}^{II}(\Omega_n^*) - T_{21}^{II}(\Omega_n^*)\right)$$
$$T_{12}^I(\Omega_n^*) + T_{21}^I(\Omega_n^*) = -T_{12}^{II}(\Omega_n^*) - T_{21}^{II}(\Omega_n^*) \qquad (32)$$

While equation (31) applies for any arbitrary frequency, the relations given in equation (32) hold only at an interface frequency, $\Omega_n^*$. It is evident from (22) that $T_{11}^I(\Omega) = T_{11}^{II}(\Omega)$, and the off-diagonal components of matrices $\boldsymbol{T}^I$ and $\boldsymbol{T}^{II}$ can be rewritten as $T_{kl}^{I(II)}(\Omega) = it_{kl}^{I(II)}(\Omega)$, where $t_{kl}^I$ and $t_{kl}^{II}$ are real numbers for $k \neq l$. By introducing these notations along with equation (32) into equation (31), one obtains the following expressions for the reflection coefficients at an interface state frequency:

$$r^I(\Omega_n^*) = \left|\frac{-i\left(t_{12}^{II}(\Omega_n^*) - t_{21}^{II}(\Omega_n^*)\right)}{2T_{11}^I(\Omega_n^*) - i\left(t_{12}^{II}(\Omega_n^*) + t_{21}^{II}(\Omega_n^*)\right)}\right|^2$$

$$r^{II}(\Omega_n^*) = \left|\frac{i\left(t_{12}^{II}(\Omega_n^*) - t_{21}^{II}(\Omega_n^*)\right)}{2T_{11}^{II}(\Omega_n^*) + i\left(t_{12}^{II}(\Omega_n^*) + t_{21}^{II}(\Omega_n^*)\right)}\right|^2 \qquad (33)$$

Interestingly, equation (33) reveals that at the interface state frequency, $\Omega_n^*$, the reflection coefficients of the two semi-infinite PCs, PC-I and PC-II, coincide, i.e., $r^I(\Omega_n^*) = r^{II}(\Omega_n^*)$. This observation can be understood through wave interference and boundary matching at the interface, which acts as a boundary where waves from each lattice meet. At the intersection of the reflection coefficients, constructive interference occurs, allowing for the formation of localized modes. This intersection also indicates an effective impedance match between the two lattices, which is crucial for the existence of the interface state, as it prevents the wave energy from fully transmitting or reflecting, thereby localizing it at the interface. To our knowledge, this specific relationship between the intersection of reflection coefficients and the precise location of interface states has not been previously reported, providing a new perspective or refinement to existing theories. Moreover, this observation provides additional insight into the existence of interface states through the reflection spectra.

## 4. Vibroacoustic topological interface states – The Lattice of finite extent

Thus far, the analysis of the topological interface states has been conducted in terms of the local transfer matrix of the composite vibroacoustic metamaterial of infinite extent composed of two semi-infinite left and right sub-lattices with different topological features. This analysis was performed independently of the number of unit-cells and without consideration of any boundary conditions. In this section, we consider the corresponding system of finite extent composed of only a finite number of coupled unit-cells. Specifically, the composite lattice described in Figure 5 is constructed by connecting two finite sub-lattices: The right sub-lattice consists of $N$ unit cells with configuration PC-I, and the left of $N$ unit cells with configuration PC-II.

By construction, both PC-I and PC-II sub-lattice configurations possess the same "spatial period" ($\Delta_1 + \Delta_2 = 2\Delta_0$), and the same propagation constant, which yields to identical band structures. However, the two sub-lattices exhibit different band topologies, as determined from the analysis of the semi-infinite lattices in the previous sections. Consequently, localized interface states are expected to emerge in the interfacial region of the finite vibroacoustic lattice as well. We aim to study the robustness of these localized interface states, which, until this point, were predicted solely based on the semi-infinite left and right lattice perspective.

The acoustic state vectors of the finite composite lattice at the left- and right-most boundaries are denoted by $\left[\frac{p_{L_1}}{\rho_a c_a} \quad v_{L_1}\right]^T$ and $\left[\frac{p_{R_{2N}}}{\rho_a c_a} \quad v_{R_{2N}}\right]^T$, respectively. Using the notation introduced in previous sections, we utilize the *local* transfer matrices $\boldsymbol{T}^I$ and $\boldsymbol{T}^{II}$, to construct the *global transfer matrix*, denoted by $\boldsymbol{T}^{tot}$, connecting state vectors at the left and right boundaries of the composite lattice as follows:

$$\begin{bmatrix} -\frac{p_{R_{2N}}}{\rho_a c_a} \\ v_{R_{2N}} \end{bmatrix} = \boldsymbol{T}^{tot} \begin{bmatrix} \frac{p_{L_1}}{\rho_a c_a} \\ v_{L_1} \end{bmatrix}, \quad \boldsymbol{T}^{tot} = (\boldsymbol{T}^I)^N (\boldsymbol{T}^{II})^N \qquad (34)$$

Based on the unimodularity of the transfer matrices $T^I$ and $T^{II}$, the $N^{th}$ powers of these matrices can be expressed, using the Cayley–Hamilton theorem, as [47],

$$(T^\#)^N = \begin{bmatrix} \dfrac{T_{11}^\# \sin(N\mu) - \sin((N-1)\mu)}{\sin(\mu)} & \dfrac{T_{12}^\# \sin(N\mu)}{\sin(\mu)} \\ \dfrac{T_{21}^\# \sin(N\mu)}{\sin(\mu)} & \dfrac{T_{22}^\# \sin(N\mu) - \sin((N-1)\mu)}{\sin(\mu)} \end{bmatrix} \quad (35)$$

where the superscript "#" can be replaced with either "I" or "II", indicating the components of matrices $T^I$ and $T^{II}$, respectively. Note, that the propagation constant, $\mu$, in (35) is identical for both configurations PC-I and PC-II. Introducing (35) into (34) and performing elementary algebraic manipulations yields the following explicit expression for the global transfer matrix, $T^{tot}$, in terms of the components of the matrices $T^I$ and $T^{II}$:

$$T^{tot} = \begin{bmatrix} T_{11}^{tot} & T_{12}^{tot} \\ T_{21}^{tot} & T_{22}^{tot} \end{bmatrix} = (T^I)^N (T^{II})^N$$

$$= \begin{bmatrix} \cos^2(N\mu) - \dfrac{T_{21}^{II}}{T_{21}^I}\sin^2(N\mu) & \dfrac{(T_{12}^I + T_{12}^{II})\sin(2N\mu)}{2\sin(\mu)} \\ \dfrac{(T_{21}^I + T_{21}^{II})\sin(2N\mu)}{2\sin(\mu)} & \cos^2(N\mu) - \dfrac{T_{12}^{II}}{T_{12}^I}\sin^2(N\mu) \end{bmatrix} \quad (36)$$

In (36), the relations $T_{11}^I = T_{22}^I = T_{11}^{II} = T_{22}^{II} = \cos(\mu)$ and the unimodularity property of the transfer matrices $T^I$ and $T^{II}$ were imposed. Unlike the local transfer matrices $T^I$ and $T^{II}$, which have identical diagonal components, the global transfer matrix $T^{tot}$ possesses distinct diagonal components, i.e., $T_{11}^{tot} \neq T_{22}^{tot}$. This difference arises due to the break of inversion symmetry in the composite lattice, as evidenced in Figure 5.

The existence of topological interface states in the finite composite lattice can be identified as peaks in the transmission spectra or dips in the reflection spectra measured at the interfacial region. To obtain the scattering (transmission and reflection) spectra for the composite lattice, we introduce the components of the global transfer matrix into the general formula for scattering spectra derived in [47]. As a result, we obtain:

$$r^{tot}(\Omega) = \left| \dfrac{T_{11}^{tot}(\Omega) + T_{12}^{tot}(\Omega) - T_{21}^{tot}(\Omega) - T_{22}^{tot}(\Omega)}{T_{11}^{tot}(\Omega) + T_{12}^{tot}(\Omega) + T_{21}^{tot}(\Omega) + T_{22}^{tot}(\Omega)} \right|^2$$

$$t^{tot}(\Omega) = \dfrac{4}{|T_{11}^{tot}(\Omega) + T_{12}^{tot}(\Omega) + T_{21}^{tot}(\Omega) + T_{22}^{tot}(\Omega)|^2} \quad (37)$$

Here $r^{tot}(\Omega)$ and $t^{tot}(\Omega)$ denote the reflection and transmission spectra of the composite lattice when excited by positive-going plane waves.

An alternative way to analyze the topological interface states in the composite lattice is through the eigenfrequency spectrum of an appropriately defined frequency response (or transmission) function, where peaks in the transmission spectra, indicate resonant transmission (i.e., perfect transparency if the peaks reach unity). By applying different boundary conditions, defined in terms of acoustic pressure and velocity, at the left and right boundaries of the finite composite vibroacoustic lattice, one can derive characteristic equations for the corresponding natural frequencies. For example, imposing free-free boundary conditions, i.e., $p_{L_1} = p_{R_{2N}} = 0$, the corresponding global transfer matrix yields the following characteristic equations:

$$F_1^{free-free} = sin(2N\mu(\Omega)) = 0$$
$$F_2^{free-free} = T_{12}^I(\Omega) + T_{12}^{II}(\Omega) = 0 \tag{38}$$

The roots of the characteristic equations provided in (38) determine the eigenfrequency spectrum corresponding to the composite lattice with both edges free. Clearly, the condition $F_1^{free-free} = 0$ can only be satisfied inside a propagation band, i.e., when $\mu$ is a real number, yielding $2N - 1$ natural frequencies within each band, representing the *bulk eigenspectrum*, of the finite free-free composite lattice. On the other hand, the condition $F_2^{free-free} = 0$ can be satisfied either at the lower boundary of the band or well within the bandgap below it. The latter scenario clearly demonstrates the existence of interface states, as it aligns with the criterion given in equation (29). Remarkably, the condition $F_2^{free-free} = 0$ implies that the existence of interface states is independent of the number of unit cells within the finite composite lattice, underscoring the robustness of these topological states.

To demonstrate the insensitivity of the bulk eigenspectrum and the topological interface states to boundary conditions, we consider a different scenario where both edges of the finite composite lattice are fixed, i.e., $v_{L_1} = v_{R_{2N}} = 0$. The corresponding frequency equation becomes:

$$F_1^{fixed-fixed} = sin(2N\mu(\Omega)) = 0$$
$$F_2^{fixed-fixed} = T_{21}^I(\Omega) + T_{21}^{II}(\Omega) = 0 \tag{39}$$

The condition $F_1^{fixed-fixed} = 0$ provides the bulk eigenspectrum when both edges of the composite lattice are fixed. It is evident that the bulk eigenspectrum (39) is identical to (38) for free-free boundary conditions. The condition $F_2^{fixed-fixed} = 0$ can be satisfied either at the upper boundary of the band or well within the bandgap above it. The second scenario is merely another version of the interface state existence criterion, as shown in (29). Therefore, both the bulk eigenspectrum and the topological interface states show robust behavior of the interface topological mode with respect to changes in the boundary conditions. The only distinction between the two boundary conditions is the occurrence of the band-edge natural frequency, which appears either at the lower or upper edge of the band.

While the insensitivity of topological interface states to the number of unit-cells within the finite composite lattice as well as to external boundary conditions is well known—largely attributed to topological protection that keeps these states localized and robust against perturbations—this has mainly been demonstrated numerically. Here, however, we systematically and rigorously prove these robust characteristics. It is worth noting that demonstrating the persistence of topological states becomes more challenging as the number of unit cells increases due to the need for extremely fine resolution. With our approach, however, this was achieved naturally.

## 5. Results and discussion

Unless stated otherwise, throughout this study the system parameters listed in Table I are considered. We used the actual values for air speed ($c_a$) and density ($\rho_a$) and selected Polydimethylsiloxane (PDMS) as the membrane material. The speed of elastic wave propagation in the membrane ($c_m$) and the depths of the cavities ($\Delta_1$ and $\Delta_2$) are treated as design parameters to explore the existence, evolution, and robustness of topological interface states in the composite multilayered vibroacoustic metamaterial, considering both infinite and finite lattice configurations.

| Parameter | Description |
| --- | --- |
| $c_a = 343 \ [m/s]$ | Speed of sound in air |
| $\rho_a = 1.255 \ [kg/m^3]$ | Air density |
| $h_m = 0.0011 \ [m]$ | Membrane thicknesses |
| $\rho_m = 1050 \ [kg/m^3]$ | Membrane density |
| $a = 0.05 \ [m]$ | Membrane radius |

**Table I**: System parameters kept fixed throughout the study.

### 5.1. Demonstrating the existence of topological interface states

Panels (a) and (b) of Fig. 6 present the six primary pass bands and bandgaps in the band structures for a vibroacoustic metamaterial with an infinite number of unit-cells, constructed uniformly with PC-I ($\gamma = 1/3$) or PC-II ($\gamma = -1/3$) configurations, respectively. For each band, the two edge (bounding) modes are classified based on the acoustic pressure distribution within a representative unit-cell as either symmetric (cyan circles) or antisymmetric (orange circles). Consequently, the Zak phase for each band, labeled in red, is determined to be $\pi$ if the band-edge states exhibit different symmetries. Conversely, if the band-edge states share the same symmetry, the Zak phase is 0.

Given that the summation of Zak phases below the second, fourth, and sixth bandgaps differs between the PC-I and PC-II configurations, edge states localized at the interface between the two PCs are expected within these bandgaps. Another indication of the non-trivial topological nature of these bandgaps is the difference in the sign of the imaginary part of $T_{21}$ within them. Therefore, according to the criterion provided in (26), these bandgaps support the formation of topological interface states. Physically, this implies that the surface impedances in PC-I and PC-II lattice configurations have opposite signs within these bandgaps, therefore preventing wave propagation across the interface and leading to the formation of a localized topological interface state.

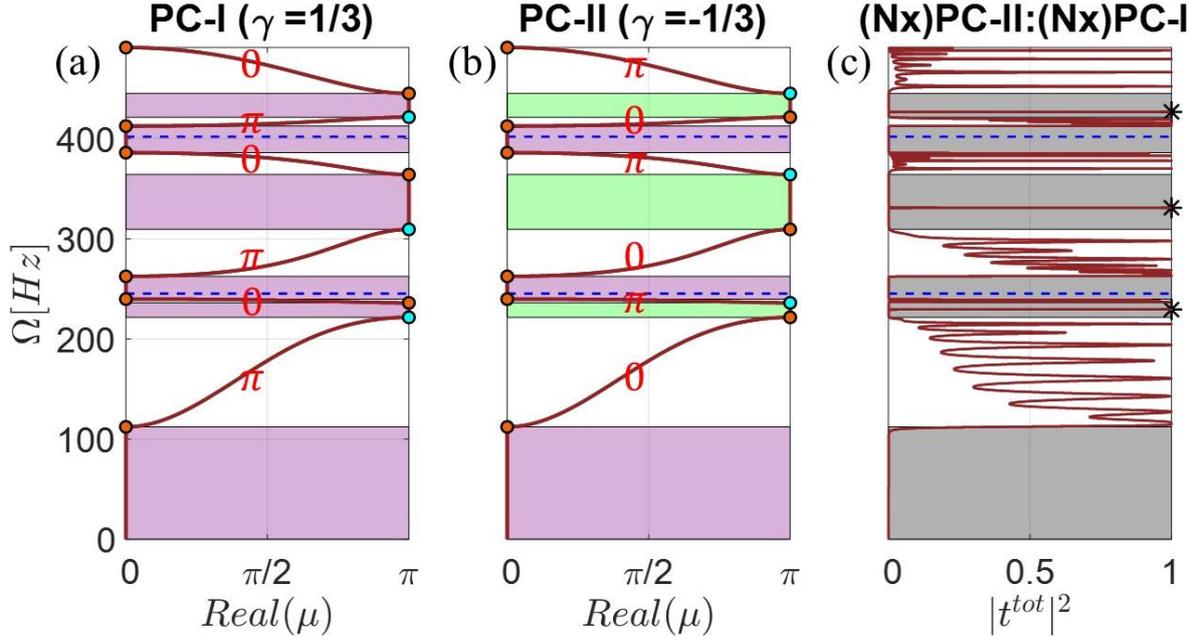

**Figure 6**: The six leading passbands and bandgaps of the band structure for a vibroacoustic metamaterial with an infinite number of uniform unit cells, with (a) PC-I ($\gamma = 1/3$), or (b) PC-II ($\gamma = -1/3$) configurations. Panel (c) shows the corresponding transmission spectrum, computed using equation (37), of a composite finite vibroacoustic lattice consisting of $N = 5$ PC-I unit-cells on the right of the interface, and $N = 5$ PC-II unit-cells on the left. System parameters are $c_m = 15[m/s]$ and $\Delta_0 = 0.06[m]$. Across the three panels, the dashed blue lines represent the two leading natural frequencies of the membrane. In panels (a) and (b), the magenta strips indicate bandgaps with $\Im(T_{21}^I) < 0$, while the green strips indicate bandgaps with $\Im(T_{21}^I) > 0$. Cyan and orange filled circles denote symmetric and antisymmetric band-edge modes, respectively, and the Zak phase of each individual band is labeled in red in (a) and (b). In panel (c), the gray strips represent bandgaps, and the black stars mark the locations of topological interface states within the topological bandgaps.

Although the existence of topological interface states was derived from the perspective of the *infinite* vibroacoustic lattices, the result in Fig. 6(c) confirms the family of topological interface modes in the composite *finite* lattice. Specifically, the transmission spectrum, computed using equation (37), for the composite finite lattice composed of a total of 10 unit-cells (5 PC-I unit-cells on the right and another 5 PC-II unit-cells on the left) confirms the topological character of the localized interface modes inside an infinite number of stopbands. The formation of these interface states, which in Figure 6(c) appear as resonant transmission peaks within the second, fourth, and sixth bandgaps, is evident and marked with black stars. The emergence of these interface states specifically within the even bandgaps, and not in others, can be attributed to the fact that as the contrast parameter $\gamma$ increases from $-1/3$, the widths of the second, fourth, and sixth bandgaps decrease, until $\gamma = 0$, where these bandgaps close, as shown in Fig. 2. With further increase of $\gamma$, these bandgaps reopen and this is accompanied by a change in the sign of the local transfer matrix component $T_{21}$ (analogous to a change in impedance sign), as well as a switch in the Zak phase of the corresponding bands. This behavior represents a topological phase transition, which occurs when two bands cross each other. Such a topological phase transition in the current vibroacoustic metamaterial is a classical analog of the SSH model in electronic systems [1-5], with the difference that in the current system there is a countable infinite of topological interface modes that are thus generated.

### 5.2. Determining the exact locations (frequencies) of topological interface states

After establishing the existence of interface states, determining their precise location within the corresponding topological bandgaps is of significant practical interest for predictive design of lattices with multi-frequency interface states. To this end, panel (a) of Fig. 7 illustrates the imaginary parts of the local transfer matrix components, $T_{21}$, for the infinite uniform vibroacoustic lattices constructed either from PC-I or PC-II unit-cells. As expected, the sign of the components $T_{21}$ remain unchanged within the bandgaps for each of these lattice configurations. Therefore, the criterion for the existence of interface states, which requires different values of $sgn(T_{21})$ for lattices composed of PC-I or PC-II unit-cells, as provided in equation (26), is absolute and functions consistently across all scenarios, with no exceptions.

Within the odd bandgaps shown in Fig. 7(a), both the PC-I (blue line) and PC-II (red line) configurations possess the same value of $sgn(T_{21})$, classifying them as topologically trivial bandgaps that do not support the formation of interface states. Conversely, the even bandgaps are topologically non-trivial, as the configurations PC-I and PC-II exhibit opposite values of $sgn(T_{21})$. Furthermore, the exact location of the interface states within these topologically non-trivial bandgaps can be exactly determined according to condition (27), i.e., at the frequency where $\Im(T_{21}^I + T_{12}^I) = 0$. The locations of the interface states within the even bandgaps correspond to the intersections of the magenta line with the zero axis, as indicated by the vertical black dashed lines. The interface topological mode locations are further confirmed by the using equation (30) yielding the same results.

Further physical insight on the interface states can be gained from the reflection coefficients of the infinite uniform lattices composed of PC-I or PC-II unit-cells. At an interface state, the reflection coefficients of these two topologically distinct lattices must intersect, as mathematically proven by equation (33). This condition is illustrated in Fig. 7(b). The presence of topological interface states within the even bandgaps is also evident from the reflection spectrum, computed using equation (37), for the composite finite vibroacoustic lattice of 10 PC-I and PC-II unit-cells, as shown in Fig. 7(c). Although the location of the interface states in the *finite* composite lattice was determined based on arguments involving the corresponding uniform infinite lattices, the formation of these localized states in the presented numerical results indicates their robustness with respect to the number of unit-cells of in the finite composite lattice, as was mathematically studied in Section 3.3.

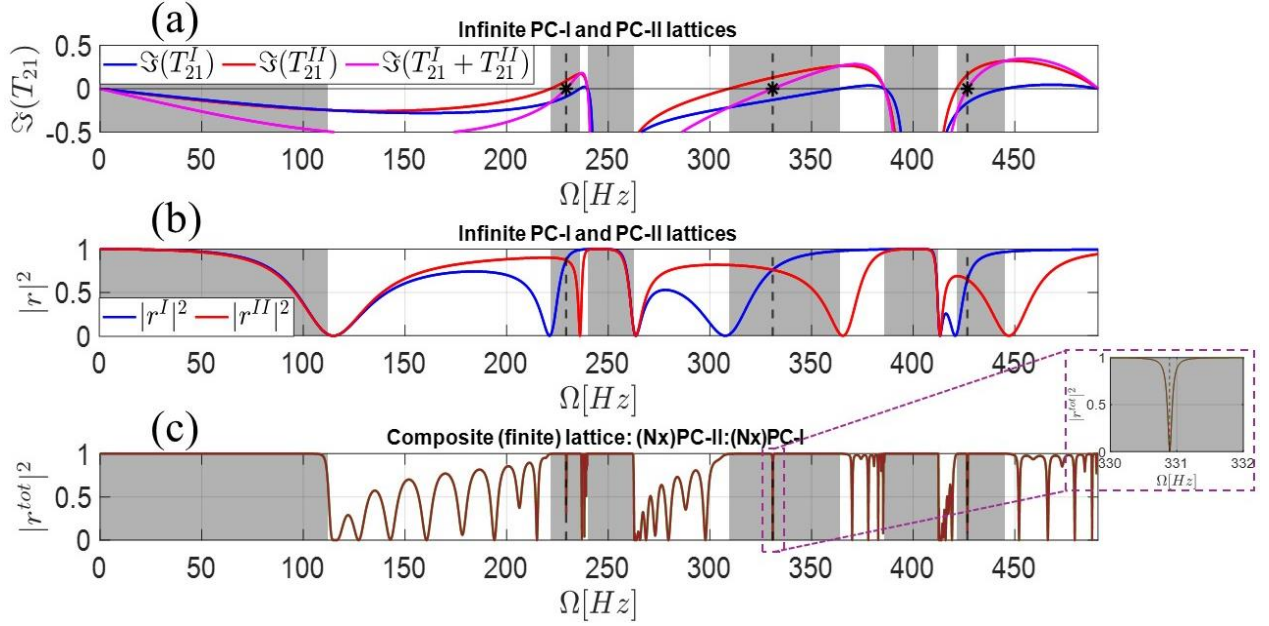

**Figure 7**: In panel (a) we depict the imaginary parts of the local transfer matrix components $T_{21}$ for infinite uniform lattices composed of PC-I unit-cells (solid blue lines), or PC-II unit-cells (solid red lines), and their summations (solid magenta lines). Panel (b) shows the corresponding reflection coefficients, computed using equation (31), using the same collor notation. Panel (c) depicts the reflection spectrum, computed using equation (37), for a composite finite lattice consisting of $N = 5$ PC-I unit cells of PC-I on the left of the interface, and $N = 5$ unit cells of PC-II unit cells on the right. System parameters are $c_m = 15[m/s]$ and $\Delta_0 = 0.06[m]$. In all panels, the dashed black columns, determined using equation (30), denote the positions of the topological interface states. The inset in panel (c) highlights the second interface state an enlargement of the reflection spectra.

### 5.3. Natural frequencies and mode shape of topological interface states

By definition, resonant transmission occurs at the natural frequencies of the finite composite vibroacoustic lattice, as given by equation (38) for free boundary conditions, or by equation (39) for fixed ones. The upper panel of Figure 8 illustrates the projection of the natural frequencies—indicated by black-edged circles—onto the third and fourth bands for a free-free multilayered vibroacoustic composite lattice consisting of 20 PC-I unit-cells on the right of the interface, and 20 PC-II unit-cells on the left. The emergence of an interface state within the fourth bandgap is evident, marked by a red-filled circle. The lower panel of Figures 8 depicts the acoustic pressure mode shape across the unit-cells for the interface mode. The strong localization of acoustic pressure at the interface between the two lattices is clearly shown.

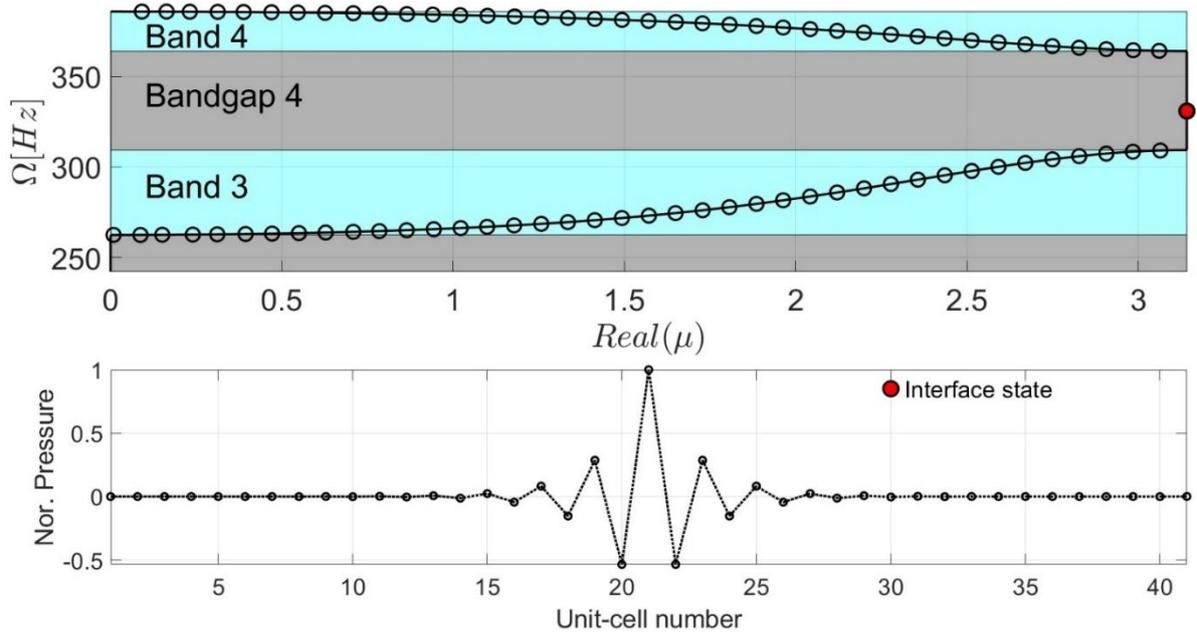

**Figure 8**: Projection of the natural frequencies (circles) for a free–free multilayered vibroacoustic composite lattice consisting of $N = 20$ PC-I unit cells on the right of the interface, and $N = 20$ PC-II unit cells on the left, onto the dispersion branches of the third and fourth passbands of the corresponding infinite composite lattice, with $c_m = 15 [m/s]$ and $\Delta_0 = 0.06 [m]$ (upper panel). Acoustic pressure mode shape for the interface mode in bandgap 4 (lower panel).

### 5.4. Evolution of topological interface states

We have mathematically proven and numerically demonstrated the insensitivity of the topological interface states (modes) in the composite vibroacoustic multilayered lattice under "perturbations" including boundary conditions, number of unit-cells (in finite systems), and variations in system parameters. Specifically, we established the independence of these states from changes in the contrast parameter $\gamma$, as described in equation (30) and illustrated in Figure 2; their independence from the number of unit cells, as shown in equation (38) and Figures 6, 7, and 8; and their insensitivity to boundary conditions, as indicated in equations (38) and (39) and depicted in Figure 8. Figures 9 and 10 further illustrate the evolution of the first three interface states as the speed of elastic wave propagation in the membrane ($c_m$) and the nominal cavity depth ($\Delta_0$) vary, respectively. It is evident that, due to the absence of band crossing events, the interface states evolve smoothly with the band-edge frequencies, consistently appearing within the Bragg-like band-splitting induced bandgaps, but not within the local resonance bandgaps. The persistence of these topological states, with the bandgaps remaining open, confirms the robustness of the topological phase against a wide range of varying system parameters. It is noteworthy that our analytical investigations have provided both straightforward and exact expressions for determining the evolution of the bands - using equation (19), and the topological interface states - using equation (30), for the considered multilayered vibroacoustic metamaterial.

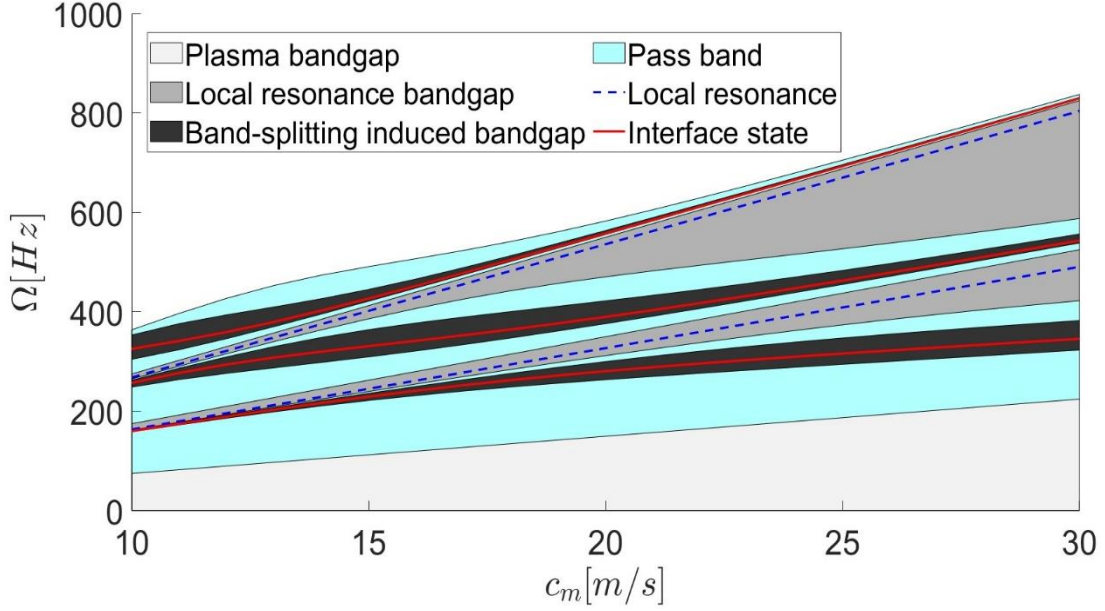

**Figure 9**: Evolution of the first three interface modes along with the corresponding band structure of the infinite multilayered vibroacoustic lattice as the nominal cavity depth is fixed to $\Delta_0 = 0.06 [m]$, while the speed of elastic wave propagation in the membrane ($c_m$) varies. The cyan strips indicate the pass bands, while the three levels of gray-scale shading, from light to dark, represent the plasma, local resonance, and band-splitting induced band gaps, respectively. The dashed blue lines denote the natural frequencies of the membrane, and the solid red lines represent the topological interface states.

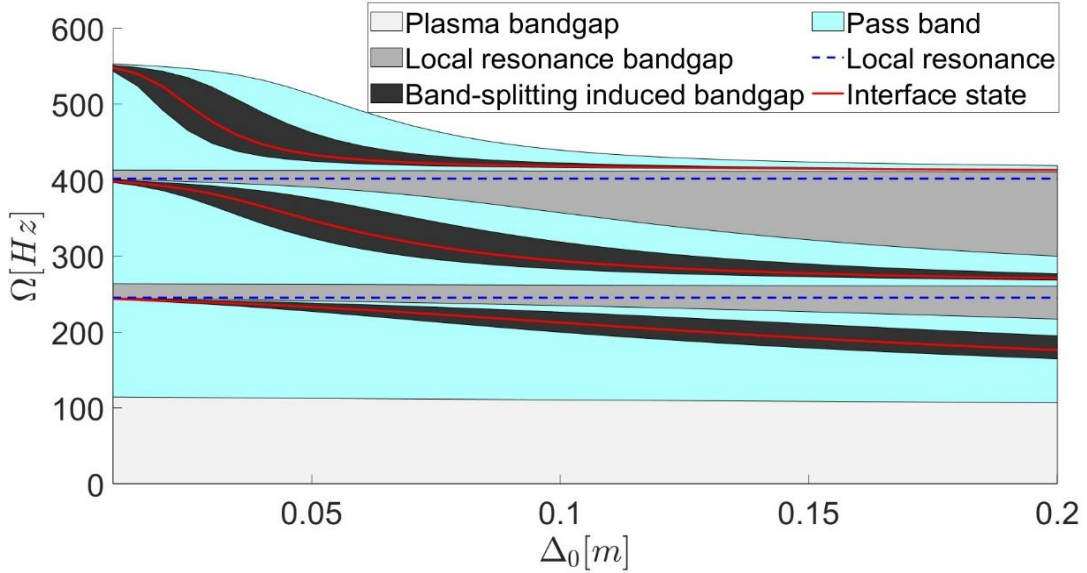

**Figure 10**: Evolution of the first three interface states along with the corresponding band structure of the infinite multilayered vibroacoustic lattice as the speed of elastic wave propagation in the membrane is fixed to $c_m = 15 [m/s]$, while the nominal cavity depth ($\Delta_0$) varies. The cyan strips indicate the pass bands, while the three levels of gray-scale shading, from light to dark, represent the plasma, local resonance, and band-splitting induced band gaps, respectively. The dashed blue lines denote the natural frequencies of the membrane, and the solid red lines represent the topological interface states.

## 5.5. Delocalization phenomenon of interface states

Fig. 11 depicts the localization behavior of the third interface state across points P1 to P6. For points P1 to P3, where the hosting bandgap is relatively wide, the interface state is strongly localized at the interfacial region. However, as the bandgap width decreases, the interface state becomes delocalized, merging with the bulk states, as shown for points P4 to P6. This hybridization results in mixed states that exhibit both strong localization at the interface and extended field distribution within the bulk. A similar delocalization behavior of edge states, accompanied by "mixed states," has been recently reported in [42, 43] while investigating the topological phases of polaritons in a cavity waveguide modeled as an extended SSH model, where a breakdown of the bulk-edge correspondence was observed.

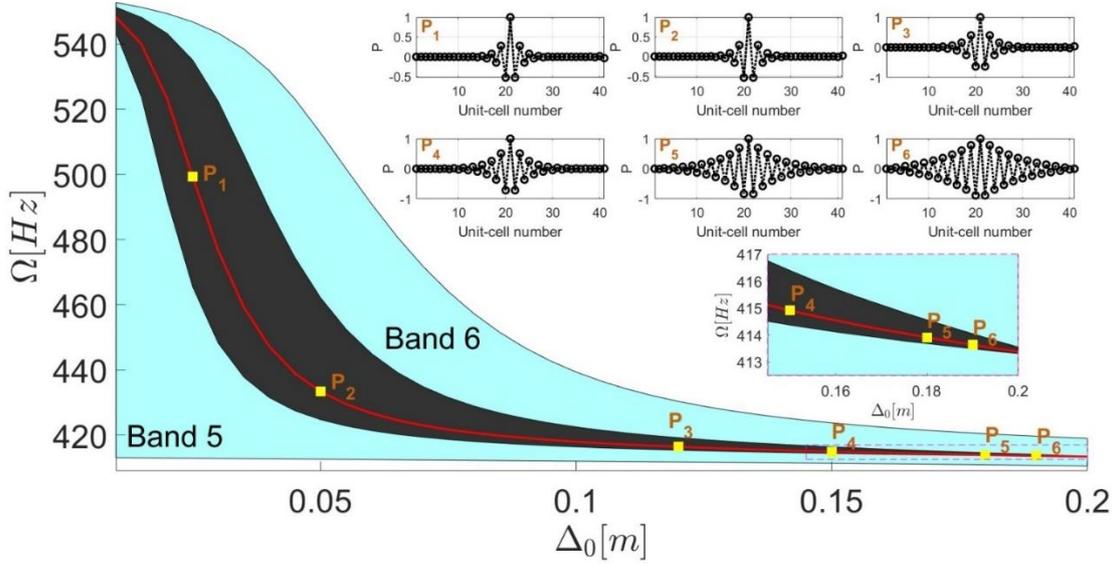

**Figure 11**: Delocalization of the third interface state, from Figure 10, in the topological phase as the nominal cavity depth ($\Delta_0$) increases. The insets show the localization of the interface state at the interfacial region along points $P1$ to $P6$. Additionally, an enlarged view of the region where the interface state hybridizes with the bulk is provided.

## 5.6. Demonstrating the robustness of topological interface states

To rigorously assess the robustness of the interface states, we introduce random variations in the membrane wave speed as follows,

$$c_m = c_0 + \delta c, \text{ where } \delta c \sim Uniform(-\alpha, \alpha) \tag{40}$$

where $\delta c$ represents the random variation uniformly distributed between $-\alpha$ and $\alpha$, with $\alpha$ denoting the magnitude of the randomness. Fig. 12 visualizes the disordered lattice and depicts the mode shape of the first interface state under both slight, intermediate, and strong perturbations. The pronounced localization of the interface states around the interfacial region underscores the resilience of these topological states to random and significant perturbations.

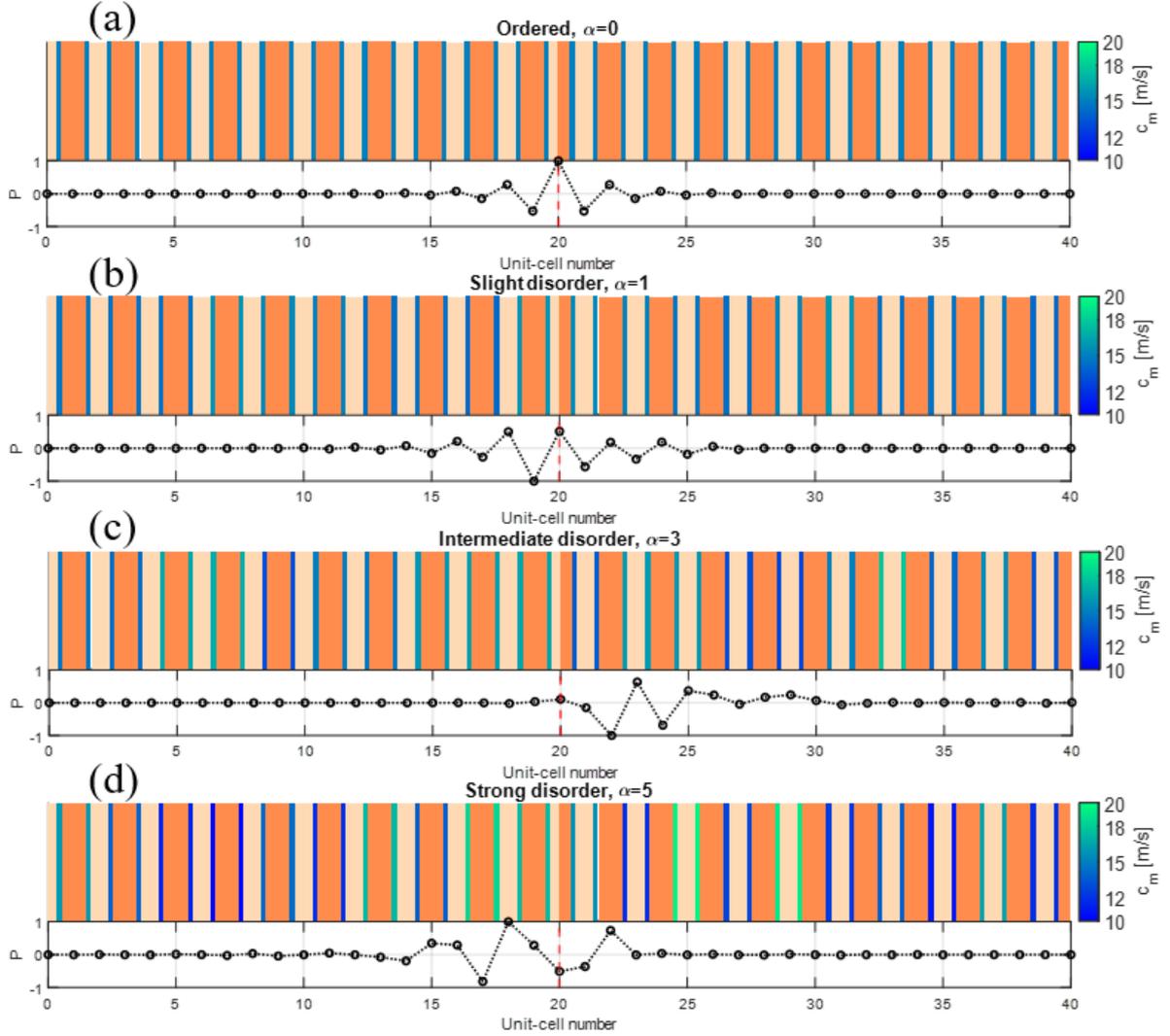

**Figure 12**: Sensitivity of interface states to varying levels of disorder in the membrane wave speed, $c_m$: ordered medium ($\alpha = 0$) panel (a), slight disorder ($\alpha = 1$) panel (b), intermediate disorder ($\alpha = 3$) panel (c), and strong disorder ($\alpha = 5$, lower panel). Light orange represents cavity 1, while dark orange represents cavity 2. The color bar indicates the magnitude of the membrane wave speed ($c_m$), which varies randomly according to Equation (40). In each panel, the mode shape corresponding to the interface state in both ordered and disordered media is depicted.

## 6. Concluding remarks

In this study, we systematically and rigorously explored the formation, evolution, and robustness of subwavelength topological interface states in a multilayered vibroacoustic phononic lattice composed of repetitive membrane-air cavity layers. Focusing on the challenging low-frequency range and assuming axisymmetric modes, we employed a purely analytical approach based on the transfer matrix formalism to analyze the interface states in inversion-symmetric unit cells. We began by solving the sound-membrane interaction problem in an exact manner within a representative unit-cell of an infinite multilayered vibroacoustic lattice, enabling a detailed characterization of its band structure. This led to the development of a systematic procedure for constructing topological interface states through phase transitions by coupling two topologically distinct semi-infinite vibroacoustic lattices. Through this framework, we derived analytical criteria for the existence of interface states based on three distinct concepts, namely, the Zak phase, surface impedance, and a newly proposed perspective based on reflection coefficients, all expressed in closed form through the components of the transfer matrix. Notably, we obtained an explicit expression for the exact location of the interface states, offering insight into their evolution and providing a practical guide for their design and realization. Furthermore, our analysis of the topological interface states within finite composite lattices demonstrated the robustness characteristic of these states against variations in the number of unit cells, as well as perturbations in system parameters and boundary conditions. Interestingly, beyond their typical strong localization at the interfacial region between two topologically distinct lattices, our findings reveal that as the width of the hosted bandgap narrows, the interface states begin to hybridize with the bulk, leading to a delocalization phenomenon. This delocalization not only offers new avenues in the design of these robust topological states but can also be harnessed to achieve tunable waveguiding and energy transfer in advanced metamaterial systems.

It is worth noting that the majority of previous studies on topological states relied heavily on finite element analysis, numerical simulations, or experimental verification to support primary analyses, often based on computational methods or analytical approximations such as infinite eigenmode expansions or effective medium theory. In contrast, our study presents an exact analysis of the interface states by directly solving the sound-membrane interaction problem in the low-frequency range, entirely free from infinite series truncations or other approximations. This precise methodology negates the need for further validation through approximated methods, highlighting the robustness and accuracy of our findings. Our work provides a comprehensive characterization of interface states purely through analysis, underscoring the strength of the transfer matrix method. Though originally derived for infinite lattices, this method enables a full characterization of interface states and offers direct physical insight into their behavior. Beyond its analytical rigor, our approach serves as a predictive design tool for practical topological lattices, paving the way for advances in applications of topological phenomena. This systematic approach is not only applicable to a broad range of vibroacoustic systems but can also be extended to other physical systems composed of repetitive unit cells with inversion symmetry, where subwavelength phenomena play a crucial role, including more general configurations of phononic and photonic crystals.

Moving forward, experimental validation will be essential for confirming the practical applicability of these results in real-world settings. These experiments will focus on verifying the robustness of the predicted topological interface states under disorder and material imperfections. Additionally, further research will explore the extension of the model to more complex multilayered systems and the incorporation of nonlinearities or non-Hermiticity to study their effects on the topological properties. Numerical simulations will complement both the analytical and experimental efforts, providing a comprehensive framework for potential practical applications in sound filters, waveguides, and acoustic sensors.

**Funding**

The authors declare financial support was received for the research and authorship of this article. The research was financially supported by the Israeli Council of Higher Education (CHE-VATAT) and the Department of Mechanical Science and Engineering (MechSE) at the University of Illinois Urbana-Champaign.

**Acknowledgments**
The authors declare that there were no contributions or discussions related to this research beyond those of the listed authors.

**Data availability statement**
The original contributions presented in the study are included in the article/Supplementary material, further inquiries can be directed to the corresponding author.

**Conflict of interest**
The authors declare that the research was conducted in the absence of any commercial or financial relationships that could be construed as a potential conflict of interest.

**Appendix**

The pressure amplitudes $P_{jc}$ and $P_{jc}$ ($j = 1,2,3$), the membranes' deflection amplitudes $W_1$ and $W_2$, and the amplitudes of the pressure and velocity at the right edge of the $n^{th}$ unit-cell $P_{R_n}$ and $V_{R_n}$, are given by:

$$P_{1c} = \cos\left(k_0 \frac{\Delta_2}{2}\right) P_{Ln} - \frac{i\sigma_a}{\Omega} \sin\left(k_0 \frac{\Delta_2}{2}\right) V_{Ln}$$
$$P_{1s} = -\sin\left(k_0 \frac{\Delta_2}{2}\right) P_{Ln} - \frac{i\sigma_a}{\Omega} \cos\left(k_0 \frac{\Delta_2}{2}\right) V_{Ln} \tag{A1}$$

$$P_{2c} = \left(\frac{\sigma_m}{\sigma_a} \sin\left(k_0 \frac{\Delta_2}{2}\right) + \cos\left(k_0 \frac{\Delta_2}{2}\right)\right) P_{Ln}$$
$$+ \frac{i\sigma_a}{\Omega} \left(\frac{\sigma_m}{\sigma_a} \cos\left(k_0 \frac{\Delta_2}{2}\right) - \sin\left(k_0 \frac{\Delta_2}{2}\right)\right) V_{Ln} \tag{A2}$$
$$P_{2s} = -\sin\left(k_0 \frac{\Delta_2}{2}\right) P_{Ln} - \frac{i\sigma_a}{\Omega} \cos\left(k_0 \frac{\Delta_2}{2}\right) V_{Ln}$$

$$P_{3c} = \left(\frac{\sigma_m}{\sigma_a}\left(\sin\left(k_0 \frac{\Delta_2}{2}\right)\cos(k_0\Delta_1) + \sin\left(k_0\left(\Delta_1 + \frac{\Delta_2}{2}\right)\right)\right)\right.$$
$$\left. + \frac{\sigma_m}{\sigma_a}\sin\left(k_0 \frac{\Delta_2}{2}\right)\sin(k_0\Delta_1)\right) + \cos\left(k_0\left(\Delta_1 + \frac{\Delta_2}{2}\right)\right) P_{Ln}$$
$$+ \frac{i}{\Omega}\left(\sigma_m\left(\cos\left(k_0 \frac{\Delta_2}{2}\right)\cos(k_0\Delta_1) + \cos\left(k_0\left(\Delta_1 + \frac{\Delta_2}{2}\right)\right)\right)\right. \tag{A3}$$
$$\left. - \sigma_a \sin\left(k_0\left(\Delta_1 + \frac{\Delta_2}{2}\right)\right) + \frac{\sigma_m^2}{\sigma_a^2}\cos\left(k_0 \frac{\Delta_2}{2}\right)\sin(k_0\Delta_1)\right) V_{Ln}$$

$$P_{3s} = -\left(\frac{\sigma_m}{\sigma_a}\sin\left(k_0\frac{\Delta_2}{2}\right)\sin(k_0\Delta_1) + \sin\left(k_0\left(\Delta_1 + \frac{\Delta_2}{2}\right)\right)\right)P_{Ln}$$
$$-\left(\frac{i\sigma_m}{\Omega}\cos\left(k_0\frac{\Delta_2}{2}\right)\sin(k_0\Delta_1) + \frac{i\sigma_a}{\Omega}\cos\left(k_0\left(\Delta_1 + \frac{\Delta_2}{2}\right)\right)\right)V_{Ln}$$

$$W_1 = \frac{1}{\sigma_a}\sin\left(k_0\frac{\Delta_2}{2}\right)P_{Ln} + \frac{i}{\Omega}\cos\left(k_0\frac{\Delta_2}{2}\right)V_{Ln}$$
$$W_2 = \left(\frac{\sigma_m}{\sigma_a^2}\sin\left(k_0\frac{\Delta_2}{2}\right)\sin(k_0\Delta_1) + \frac{1}{\sigma_a}\sin\left(k_0\left(\Delta_1 + \frac{\Delta_2}{2}\right)\right)\right)P_{Ln} \quad\text{(A4)}$$
$$+ \frac{i}{\Omega}\left(\frac{\sigma_m}{\sigma_a}\cos\left(k_0\frac{\Delta_2}{2}\right)\sin(k_0\Delta_1) + \cos\left(k_0\left(\Delta_1 + \frac{\Delta_2}{2}\right)\right)\right)V_{Ln}$$

$$P_{R_n} = -\left(\cos(k_0(\Delta_1 + \Delta_2)) + \frac{\sigma_m}{\sigma_a}\sin(k_0(\Delta_1 + \Delta_2))\right.$$
$$\left. + \frac{\sigma_m^2}{2\sigma_a^2}\sin(k_0\Delta_1)\sin(k_0\Delta_2)\right)P_{Ln}$$
$$-\frac{i}{\Omega}\left(2\sigma_m\cos\left(k_0\left(\Delta_1 + \frac{\Delta_2}{2}\right)\right)\cos\left(k_0\frac{\Delta_2}{2}\right)\right.$$
$$\left. - \sigma_a\sin(k_0(\Delta_1 + \Delta_2)) + \frac{\sigma_m^2}{\sigma_a}\sin(k_0\Delta_1)\cos^2\left(k_0\frac{\Delta_2}{2}\right)\right)V_{Ln}$$
$$v_R^{(n)}(t) = V_{R_n}e^{i\Omega t} \quad\text{(A5)}$$
$$V_{R_n} = \frac{\Omega}{i\sigma_a}\left(\sin(k_0(\Delta_1 + \Delta_2)) + 2\frac{\sigma_m}{\sigma_a}\sin\left(k_0\left(\Delta_1 + \frac{\Delta_2}{2}\right)\right)\sin\left(k_0\frac{\Delta_2}{2}\right)\right.$$
$$\left. + \frac{\sigma_m^2}{\sigma_a^2}\sin(k_0\Delta_1)\sin^2\left(k_0\frac{\Delta_2}{2}\right)\right)P_{Ln}$$
$$+ \left(\frac{\sigma_m}{\sigma_a}\sin(k_0(\Delta_1 + \Delta_2)) + \cos(k_0(\Delta_1 + \Delta_2))\right.$$
$$\left. + \frac{\sigma_m^2}{2\sigma_a^2}\sin(k_0\Delta_1)\sin(k_0\Delta_2)\right)V_{Ln}$$